\documentclass[a4paper,12pt]{article}
\usepackage{amsfonts,amsthm,amsmath,amssymb,graphicx}
\usepackage{braket}
\usepackage{color}
\usepackage{hyperref}
\begin{document}

\newtheorem{definition}{Definition}[section]
\newcommand{\be}{\begin{equation}}
\newcommand{\ee}{\end{equation}}
\newcommand{\bea}{\begin{eqnarray}}
\newcommand{\eea}{\end{eqnarray}}
\newcommand{\LE}{\left[}
\newcommand{\R}{\right]}
\newcommand{\nn}{\nonumber}
\newcommand{\Tr}{\text{Tr}}
\newcommand{\N}{\mathcal{N}}
\newcommand{\G}{\Gamma}
\newcommand{\vf}{\varphi}
\newcommand{\LL}{\mathcal{L}}
\newcommand{\Op}{\mathcal{O}}
\newcommand{\HH}{\mathcal{H}}
\newcommand{\arctanh}{\text{arctanh}}
\newcommand{\up}{\uparrow}
\newcommand{\down}{\downarrow}
\newcommand{\rd}{\partial}
\newcommand{\de}{\partial}
\newcommand{\ba}{\begin{eqnarray}}
\newcommand{\ea}{\end{eqnarray}}
\newcommand{\db}{\bar{\partial}}
\newcommand{\we}{\wedge}
\newcommand{\ca}{\mathcal}
\newcommand{\lr}{\leftrightarrow}
\newcommand{\f}{\frac}
\newcommand{\s}{\sqrt}
\newcommand{\vp}{\varphi}
\newcommand{\hvp}{\hat{\varphi}}
\newcommand{\tvp}{\tilde{\varphi}}
\newcommand{\tp}{\tilde{\phi}}
\newcommand{\ti}{\tilde}
\newcommand{\ap}{\alpha}
\newcommand{\pr}{\propto}
\newcommand{\mb}{\mathbf}
\newcommand{\ddd}{\cdot\cdot\cdot}
\newcommand{\no}{\nonumber \\}
\newcommand{\la}{\langle}
\newcommand{\lb}{\rangle}
\newcommand{\ep}{\epsilon}
 \def\we{\wedge}
 \def\lr{\leftrightarrow}
 \def\f {\frac}
 \def\ti{\tilde}
 \def\ap{\alpha}
 \def\pr{\propto}
 \def\mb{\mathbf}
 \def\ddd{\cdot\cdot\cdot}
 \def\no{\nonumber \\}
 \def\la{\langle}
 \def\lb{\rangle}
 \def\ep{\epsilon}
\def\m{{\mu}}
 \def\w{{\omega}}
 \def\n{{\nu}}
 \def\ep{{\epsilon}}
 \def\d{{\delta}}
 \def\rh{\rho}
 \def\t{{\theta}}
 \def\a{{\alpha}}
 \def\T{{\Theta}}
 \def\frac#1#2{{#1\over #2}}
 \def\l{{\lambda}}
 \def\G{{\Gamma}}
 \def\D{{\Delta}}
 \def\g{{\gamma}}
 \def\s{\sqrt}
 \def\ch{{\chi}}
 \def\b{{\beta}}
 \def\CA{{\cal A}}
 \def\CC{{\cal C}}
 \def\CI{{\cal I}}
 \def\CO{{\cal O}}
 \def\o{{\rm ord}}
 \def\Ph{{\Phi }}
 \def\L{{\Lambda}}
 \def\CN{{\cal N}}
 \def\p{\partial}
 \def\pslash{\p \llap{/}}
 \def\Dslash{D \llap{/}}
 \def\Mp{m_{{\rm P}}}
 \def\apm{{\alpha'}}
 \def\r{\rightarrow}
 \def\Re{{\rm Re}}
 \def\MG{{\bf MG:}}
\def\be{\begin{equation}}
\def\ee{\end{equation}}
\def\ba{\begin{eqnarray}}
\def\ea{\end{eqnarray}}
\def\bal{\begin{align}}
\def\eal{\end{align}}

 \def\de{\partial}
 \def\db{\bar{\partial}}
 \def\we{\wedge}
 \def\lr{\leftrightarrow}
 \def\f {\frac}
 \def\ti{\tilde}
 \def\ap{\alpha}
  \def\al{\alpha'}
 \def\pr{\propto}
 \def\mb{\mathbf}
 \def\ddd{\cdot\cdot\cdot}
 \def\no{\nonumber \\}
\def\nn{\nonumber \\}
 \def\la{\langle}
 \def\lb{\rangle}
 \def\ep{\epsilon}
 \def\ddbp{\mbox{D}p-\overline{\mbox{D}p}}
 \def\ddbt{\mbox{D}2-\overline{\mbox{D}2}}
 \def\ov{\overline}
 \def\cl{\centerline}
 \def\vp{\varphi}
\def\hB{\hat \Box}

\begin{titlepage}

\thispagestyle{empty}

\begin{flushright}
YITP-18-12\\
IPMU18-0035\\
\end{flushright}

\vspace{.4cm}
\begin{center}
\noindent{\large \bf Entanglement of Purification in Free Scalar Field Theories}\\
\vspace{2cm}

Arpan Bhattacharyya$^{a}$, Tadashi Takayanagi$^{a,b}$ and Koji Umemoto$^{a}$
\vspace{1cm}

{\it
$^{a}$Center for Gravitational Physics, \\
Yukawa Institute for Theoretical Physics (YITP), Kyoto University, \\
Kitashirakawa Oiwakecho, Sakyo-ku, Kyoto 606-8502, Japan\\ \vspace{3mm}
$^{b}$Kavli Institute for the Physics and Mathematics of the Universe,\\
University of Tokyo, Kashiwano-ha, Kashiwa, Chiba 277-8582, Japan\\
}

\vskip 2em
\end{center}

\vspace{.5cm}
\begin{abstract}

We compute the entanglement of purification (EoP) in a 2d free scalar field theory with various masses. This quantity measures correlations between two subsystems and is reduced to the entanglement entropy when the total system is pure. We obtain explicit numerical values by assuming minimal gaussian wave functionals for the purified states. We find that when the distance between the subsystems is large, the EoP behaves like the mutual information. However, when the distance is small, the EoP shows a characteristic behavior which qualitatively agrees with the conjectured holographic computation and which is different from that of the mutual information. We also study behaviors of mutual information in purified spaces and violations of monogamy/strong superadditivity.

\end{abstract}

\end{titlepage}

\newpage

\section{Introduction}

The entanglement entropy has played important roles to uncover dynamical aspects of
not only quantum field theories \cite{BKLS,Sr,HLW,CC,CHR} but also gravitational physics through holography \cite{Ma,RT}. An ideal measure of correlation between two subsystems $A$ and $B$ is the entanglement entropy $S_A(=S_B)$ if the total system $AB$ is a pure state $|\Psi\lb_{AB}$. Moreover, it coincides with the amount of quantum entanglement based on an operational viewpoint of LOCC (local operations and classical communication) \cite{NC}. For a review of the entanglement measures, refer to e.g.\cite{HHHH,Book}. This quantity is defined by the von Neumann entropy $S_A=-\mbox{Tr}[\rho_A\log\rho_A]$ of the reduced density matrix $\rho_A=\mbox{Tr}_B|\Psi\lb\la\Psi|$ for the subsystem $A$.

When the total system $AB$ is described by a mixed state $\rho_{AB}$, the entanglement entropy itself is no longer a correlation measure (for example, in general, we have $S_A\neq S_B$). In this case, there are many known correlation measures denoted as $E_{\#}(\rho_{AB})$. The most tractable quantity is the mutual information $I(A:B)=S_A+S_B-S_{AB}$.  Computations of mutual information are clearly as easy as those of entanglement entropy and have been performed by many authors.

 Another interesting correlation measure is the entanglement of purification (EoP), which is written as $E_P(\rho_{AB})$, first introduced in \cite{EP}. By purifying the mixed state $\rho_{AB}$ in a larger system $AB\ti{A}\ti{B}$, this quantity $E_P(\rho_{AB})$ is defined by the minimum of the entanglement entropy $S_{A\ti{A}}$ against all possible purifications.
 As is obvious from this definition, when the total system $AB$ is pure, $E_P(\rho_{AB})$ just coincides with the entanglement entropy $S_A(=S_B)$. In this sense, we can regard the EoP as a generalization of entanglement entropy to mixed states. It is also worth mentioning that the EoP has an interesting operational interpretation in terms of LOq (local operations and a small amount of communication).

Recently, a holographic formula for the EoP has been proposed in \cite{TaUm,EoPS}
(refer to \cite{Bao} for its generalization). The holographic EoP is given by the minimal cross-section of entanglement wedge \cite{EW1,EW2,EW3} and non-trivially satisfies the basic properties of EoP \cite{EP,Ba}. When the total system $AB$ is pure, then the holographic EoP is reduced to the holographic entanglement entropy \cite{RT} as expected.

Motivated by the simple holographic interpretation and by the interest from quantum information-theoretic viewpoints, the purpose of the present paper is to explore calculations of EoP in quantum field theories. In earlier works \cite{EoPS,EoPC,EoPT}, the EoP was computed numerically in spin systems assuming tensor network ansatz. In our paper, we would like to numerically study a free scalar field theory with a lattice discretization as was done in the very first studies of entanglement entropy \cite{BKLS,Sr}. We will focus on the ground state of a free scalar field theory in $1+1$ dimension.

An important and new feature of the EoP calculations is that we need to minimize the entanglement entropy against all possible purifications. At first sight, this looks almost impossible. To overcome this problem, we make a crucial assumption that we can restrict to gaussian wave functionals with minimal sizes in this purification procedure. This allows us to explicitly figure out the numerical values of EoP. As we will explain below there are numerical evidences that our ansatz might not be an approximation but also an exact answer. However if without this argument, our numerical results can at least serve as upper bounds of the correct EoP values.

We have to admit the fact that neither the EoP nor mutual information is appropriate measures of quantum entanglement between $A$ and $B$. This is because they are not monotonically decreasing under LOCC. In fact, several quantities, such as entanglement of formation~\cite{EoF}, relative entropy of entanglement~\cite{REE} and squashed entanglement~\cite{Tu,Esq} etc., have been defined and known to satisfy\footnote{The quantity called negativity \cite{Neg,CCT} is also an interesting correlation measure between two subsystems, which does not involve minimization procedures. However, this quantity does not satisfy all properties required for entanglement measures. Also, it does not coincide with the (von Neumann) entanglement entropy when the total system is pure. Moreover, it is natural to expect that this quantity will not have a simple holographic dual in terms of a tractable geometric quantity in generic setups, especially in higher dimensions. This is partly because it coincides not with
the von Neumann ($n=1$) but the R{\'e}nyi entropy at $n=1/2$ when the system is pure. See also recent discussions in e.g.\cite{HNeg}.} the basic properties of entanglement measures for mixed states (see reviews \cite{HHHH,Book}). However, they always involve minimization procedures, which are more complicated than the one for the EoP (refer to \cite{EoFG} for a Gaussian ansatz for entanglement of formation). For computational difficulity of entanglement measures refer to \cite{YH}, where it has been established that they are NP-hard justifying that EoP might be a good starting point even if it is not an entanglement measure. In this sense, our analysis of EoP can be regarded as the first step toward computations of entanglement measures of mixed states in field theories.

This paper is organized as follows. In section two, we will briefly review the definition and properties of entanglement of purification (EoP) as well as its holographic counterpart.
In section three, we present our general strategy to numerically calculate the EoP in a free scalar field theory. In section four, we will provide explicit numerical results of EoP. In section five, we will study the behaviors of mutual information between various subsystems. In section six, we will examine whether inequalities of monogamy and strong superadditivity are satisfied or not in our examples. In section seven we summarize our conclusions and discuss future problems.

\section{Entanglement of Purification and Holographic Dual}
In this section, we briefly review the basics of the entanglement of purification, which include the definition and information-theoretic properties of EoP. We also give a summary of the recently conjectured holographic computation of EoP and its implications.

\subsection{Definition of Entanglement of Purification and Its properties}

Let us consider a mixed state $\rho_{AB}$ in a bipartite system $AB$.
We can always purify this mixed state by extending the Hilbert space
from $\mathcal{H}_{A}\otimes\mathcal{H}_{B}$ to $\mathcal{H}_{A}\otimes\mathcal{H}_{B}\otimes\mathcal{H}_{\tilde{A}}\otimes\mathcal{H}_{\tilde{B}}$
such that the total state $\rho_{A\tilde{A}B\tilde{B}}$ is pure and
$\rho_{AB}$ is embedded in it:
\begin{equation}
\rho_{A\tilde{A}B\tilde{B}}=\ket{\Psi_{A\tilde{A}B\tilde{B}}}\bra{\Psi_{A\tilde{A}B\tilde{B}}},\ {\rm Tr}_{\tilde{A}\tilde{B}}[\rho_{A\tilde{A}B\tilde{B}}]=\rho_{AB}.
\end{equation}
Such a pure state $\ket{\Psi_{A\tilde{A}B\tilde{B}}}$ is called a purification
of $\rho_{AB}$. Note that a purification of a given state $\rho_{AB}$
is not unique and in general there are infinitely many ways to purify it.

The entanglement of purification (EoP) of $\rho_{AB}$ is defined
by minimizing the entanglement entropy $S_{A\tilde{A}}(=S_{B\tilde{B}})$
over all possible purifications of $\rho_{AB}$ \cite{EP}:
\begin{equation}
E_{P}(\rho_{AB})\equiv\min_{\ket{\Psi_{A\tilde{A}B\tilde{B}}}:{\rm purifications\ of\ }\rho_{AB}}S_{A\tilde{A}}.
\label{EoP}
\end{equation}
Here $S_{A\tilde{A}}$ is the von Neumann entropy of $\rho_{A\tilde{A}}={\rm Tr}_{B\tilde{B}}[\ket{\Psi_{A\tilde{A}B\tilde{B}}}\bra{\Psi_{A\tilde{A}B\tilde{B}}}]$.
Thus the EoP can be understood as a minimal amount of quantum entanglement
between $A\tilde{A}$ and $B\tilde{B}$ in the extended system.

The general properties of EoP are intensively studied in \cite{Ba} (See also \cite{EP,CW}). We briefly review a part of them. First,
as we already noted in the introduction, the EoP itself is not just a measure
of quantum entanglement between $A$ and $B$, but is a measure of both classical/quantum
correlations between them. In other words, EoP always vanishes
for all product states ($\rho_{AB}=\rho_{A}\otimes\rho_{B}$) and
is strictly positive for any non-product states. Moreover, $E_{P}(\rho_{AB})$
coincides with the entanglement entropy $S_{A}(=S_{B})$ when $\rho_{AB}$
is pure (i.e. when there is no classical correlation between $A$
and $B$). This fact allow us to regard the EoP as a generalization of
the entanglement entropy to a  measure of correlation for mixed states.

There are several inequalities that the EoP enjoys. For instance, the EoP
is always bounded from above by the von Neumann entropies, and from
below by a half of the mutual information:
\begin{equation}
\frac{I(A:B)}{2}\leq E_{P}(A:B)\leq\min\{S_{A},S_{B}\}.
\end{equation}
Here we simply write $E_{P}(A:B)\equiv E_{P}(\rho_{AB})$. Similarly, the EoP
satisfies the following inequality for all tripartite states $\rho_{AB_{1}B_{2}}$:
\begin{equation}
\frac{I(A:B_{1})+I(A:B_{2})}{2}\leq E_{P}(A:B_{1}B_{2}).
\end{equation}
The mutual informations on the left-hand side are based on the reduced
density matrices $\rho_{AB_{1}}={\rm Tr}_{B_{2}}[\rho_{AB_{1}B_{2}}]$ and $\rho_{AB_{2}}={\rm Tr}_{B_{1}}[\rho_{AB_{2}B_{1}}]$. The EoP on the right-hand side measures the correlation between
$A$ and $B_{1}B_{2}$.

In particular, if the $\rho_{AB_{1}B_{2}}$ is pure, the EoP  satisfies
the polygamy inequality:
\begin{equation} \label{Plg}
E_{P}(A:B_{1}B_{2})\leq E_{P}(A:B_{1})+E_{P}(A:B_{2}).
\end{equation}
On the other hand, the reverse of (\ref{Plg}) is called monogamy and the EoP sometimes satisfies this for mixed states\footnote{Only special entanglement measures, such as the squashed entanglement, can always satisfy the monogamy~\cite{KW}.}. We will discuss this more in section~(\ref{VioMonoSSA}).

Furthermore, as expected to be true for any correlation measures, the EoP never increases
upon discarding ancilla for any states (sometime called as extensivity):
\begin{equation}
E_{P}(A:B_{1}B_{2})\geq E_{P}(A:B_{1}).
\label{NIuDA}
\end{equation}

\subsection{Holographic Entanglement of Purification}\label{Holpuri}

\begin{figure}
\centering{}\includegraphics[width=0.3 \textwidth]{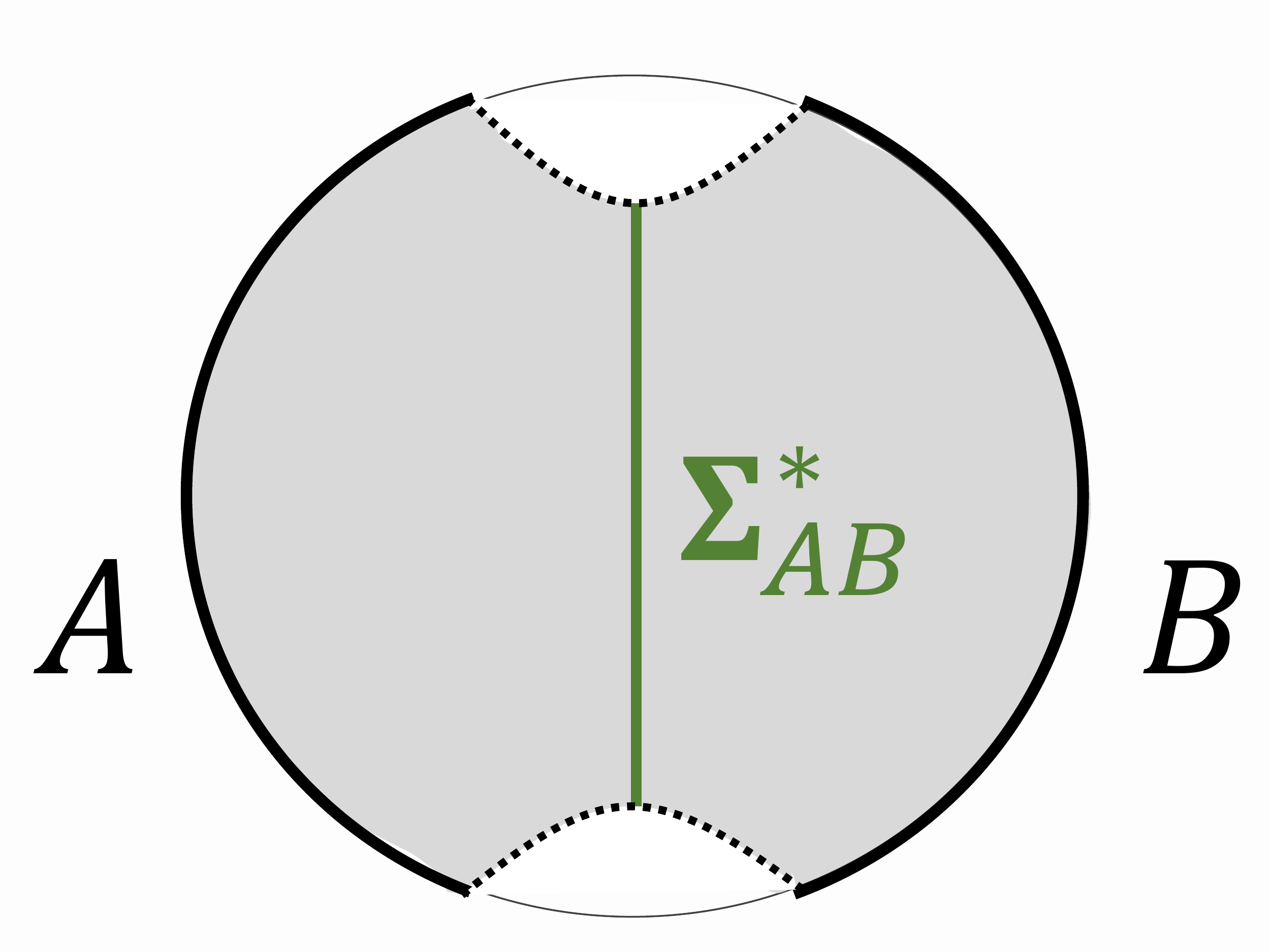}
\caption{Holographic entanglement of purification. The shaded
region is the entanglement wedge of the subsystems $A$ and $B$ in
holographic CFTs (we take a constant time slice of global AdS). The
dotted lines are the minimal surface whose area gives $S_{AB}$. The entanglement
wedge cross-section $E_{W}(A:B)$ is defined by the minimal area (divided
by $4G_{N}$) of codimension-2 surfaces which divide the entanglement
wedge into two parts. In this figure this minimal surface is denoted
by $\Sigma_{AB}^{*}$ and $E_{W}(A:B)=\frac{{\rm Area}(\sum_{AB}^{*})}{4G_{N}}$. }
\label{fig:HEoP}
\end{figure}

In \cite{TaUm,EoPS} the holographic counterpart of EoP was proposed
in the context of the AdS/CFT correspondence in the classical gravity
limit. This is the entanglement wedge cross-section denoted by $E_{W}(\rho_{AB})$. It represents
the minimal cross-section of entanglement wedge \cite{EW1,EW2,EW3} in the bulk AdS spacetime,  refer to Fig.~(\ref{fig:HEoP}).
This gives a generalization of the holographic formula of entanglement
entropy \cite{RT}. This $E_{P}=E_{W}$ (or holographic entanglement
of purification) conjecture is supported by many facts, including
the coincidence of all properties discussed in the previous section,
as well as the heuristic derivation based on the tensor network description
of the AdS/CFT correspondence. It has also an interesting connection to the bit threads picture
\cite{FH}. A generalization
of this conjecture was also discussed in \cite{Bao} and the
results further support it.

\begin{figure}[htbp]
 \begin{minipage}{0.5\hsize}
  \begin{center}
   \includegraphics[width=70mm]{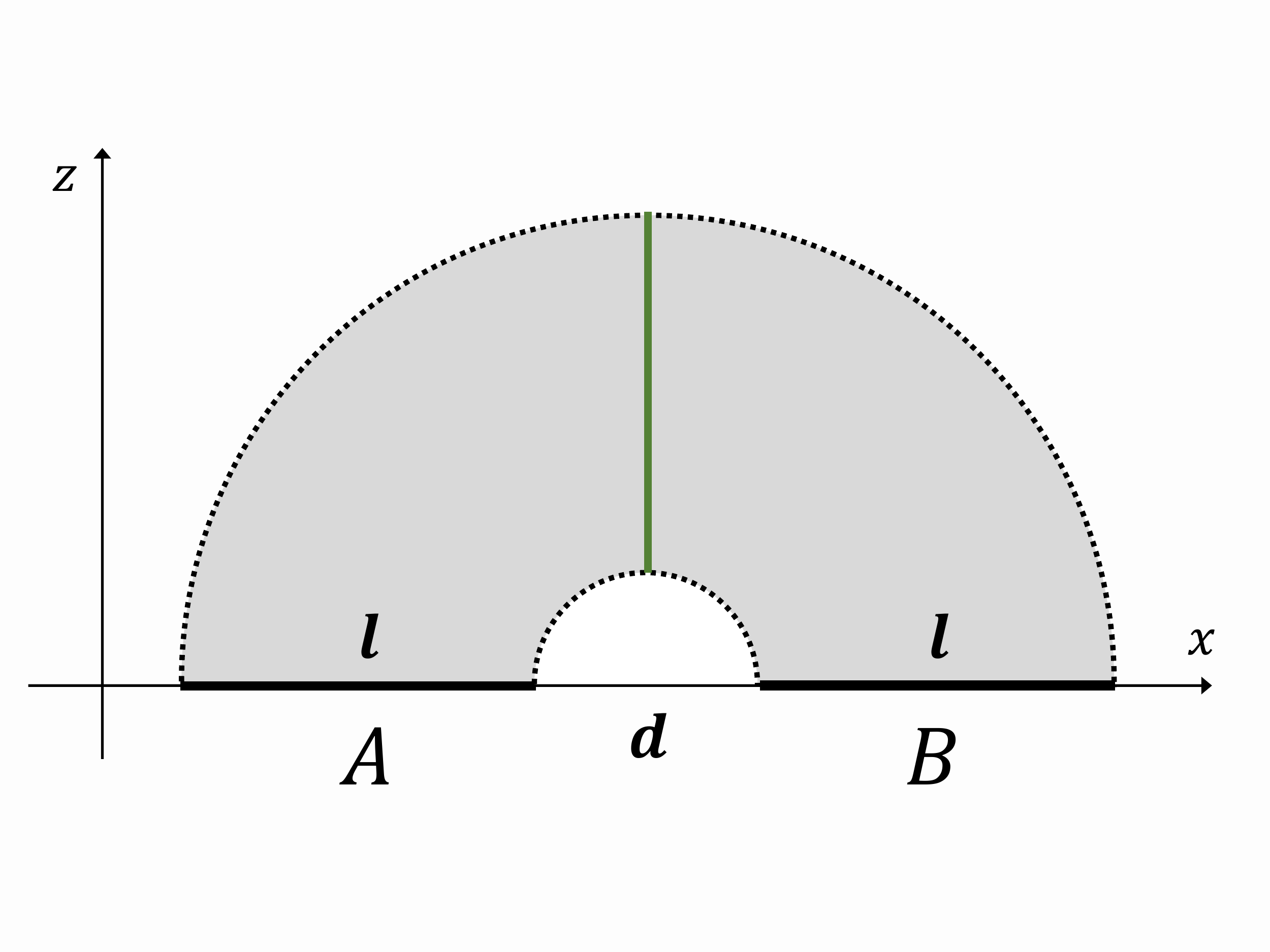}
  \end{center}
 \end{minipage}
 \begin{minipage}{0.5\hsize}
  \begin{center}
   \includegraphics[width=70mm]{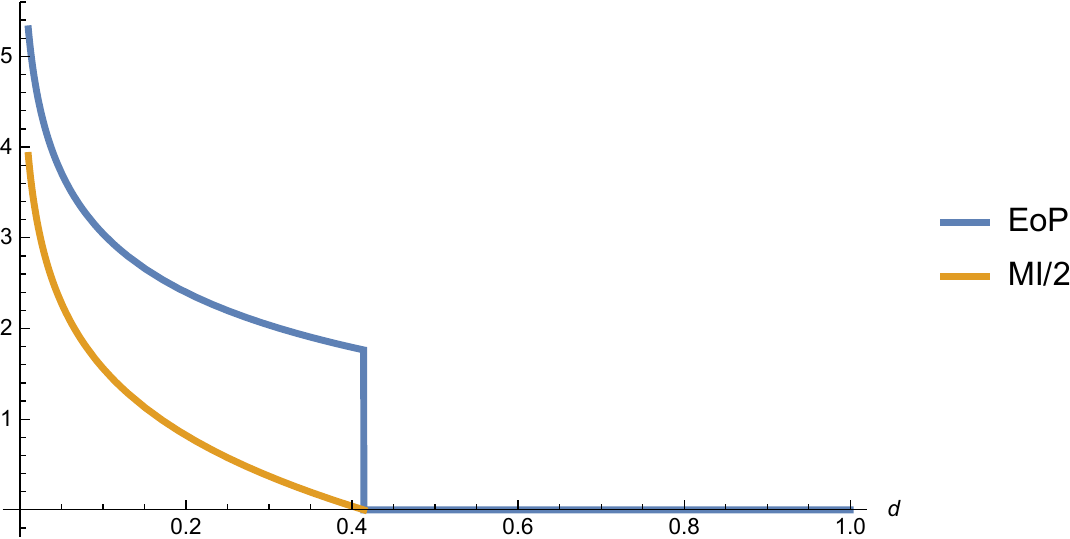}
  \end{center}
 \end{minipage}
\caption{The setup for the computation of the holographic EoP $E_{W}(A:B)$ in Poincar{\'e} AdS$_3$
(the left picture), and the plots of $E_{W}(A:B)$ (the blue curve in the right picture)
and half of holographic mutual information $I(A:B)$ (the orange curve in the right picture)
as the functions of the distance ($d$) between $A$ and $B$. Both holographic EoP and mutual information show phase transition behaviors, though only the EoP is discontinuous. We set $\frac{c}{6}=1$ and
the size $l=1$ with the transition point $d_{*}=\sqrt{2}-1$. After
the phase transition, EoP and mutual information become zero.}
\label{fig:PhaseTransition}
\end{figure}

A phase transition occurs for the holographic entanglement of purification when we change the distance between
$A$ and $B$ in holographic states. For example, in the Poincar\'e
 AdS$_{3}$ geometry which is dual to a 2d CFT on an infinite space, $E_{W}(A:B)$ can be explicitly written as
\begin{equation}
E_{W}(A:B)=\begin{cases}
\frac{c}{6}\log\left[1+\frac{2l}{d}\right], & d<(\sqrt{2}-1)l,\\
0 & d>(\sqrt{2}-1)l,
\end{cases}
\end{equation}
where $c$ is the central charge of 2d CFT and $d$ is the distance between
$A$ and $B$. We set  both the sizes of $A$ and $B$ to be $l$ for
simplicity. At the transition $d_{*}=(\sqrt{2}-1)l$ the value of of the EoP jumps, thereby providing a non-zero gap: $\Delta E_{W}=\frac{c}{6}\log[3+2\sqrt{2}]$.
We plot a typical behavior near to the transition point in Fig.~(\ref{fig:PhaseTransition}).
Mutual information $I(A:B)$ also exhibits a phase transition \cite{HRe} 
at the same point $d_{*}$ as described in the Fig.~(\ref{fig:PhaseTransition}). However, unlike EoP, the mutual information
smoothly goes to zero.

\begin{figure}[h!]
\centering{}\includegraphics[width=6.0cm]{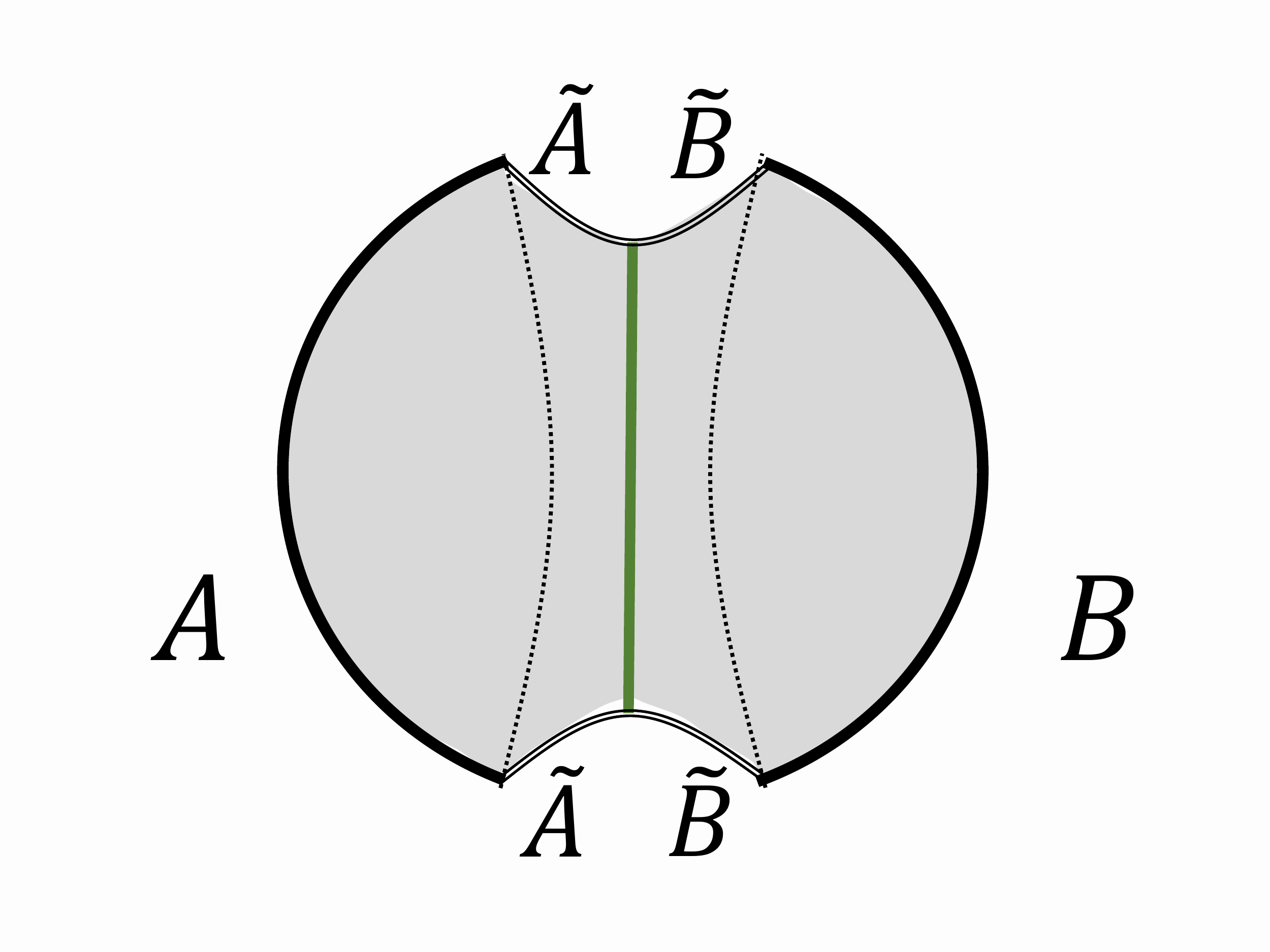}\caption{A derivation of $E_{P}=E_{W}$ based on the tensor
network description of AdS space. We regard $A\tilde{A}B\tilde{B}$
as a new boundary of bulk spacetime defining an extended field theory.
The subsystems $\tilde{A}$ and $\tilde{B},$ lying on the minimal surface used for computing $S_{AB},$ are
identified with the ancilla system. The dashed lines denotes the minimal surfaces whose areas give $S_A$ or $S_B$, respectively. Now we have to minimize the $S_{A\tilde A}$ and that is achieved by minimizing the cross-section of the wedge and that surface is denoted by the thick green line. }
\label{fig:HEoPTN}
\end{figure}

The tensor network description \cite{TNa,TNb,TNc,TNd} and the surface/state correspondence 
\cite{MiTa} give us a heuristic understanding
why $E_{P}=E_{W}$ holds \cite{TaUm}. Refer to Fig.~(\ref{fig:HEoPTN}).

\par

\begin{figure}[h!]
\centering{}\includegraphics[scale=0.75]{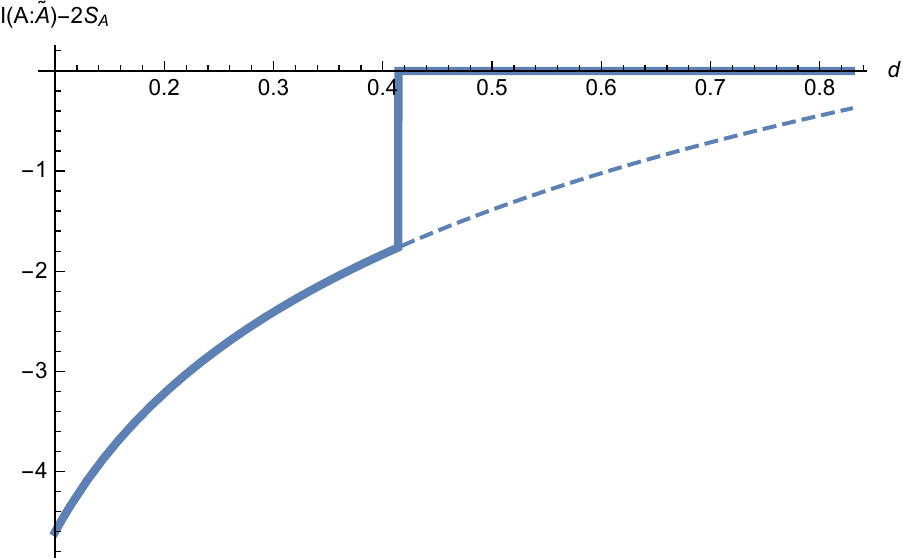}
\caption{The holographic mutual information $I(A:\tilde{A})$ subtracted
by $2S_{A}$. We set $\frac{c}{6}=1$
and the size $l=1$. It monotonically increases if we do not care
about the phase transition at $d_{*}=\sqrt{2}-1$.}
\label{fig:MIAtildeA}
\end{figure}

It also allows us to read off the properties of the mutual informations for $A,\ B,\ \tilde{A},\ \tilde{B}$. Let us consider them assuming a non-trivial situation $E_{W}(A:B)>0$.
First, we observe that $S_{\tilde{A}}$ is the area of $\tilde{A}$
itself\footnote{The reader may worry about another possible choice of the minimal surface of $S_{\tilde{A}}$ which
leads $S_{\tilde{A}}\geq S_{A}+E_{W}(A:B)$. However, in such a case we
always have the disjointed entanglement wedge, as easily shown by $I(A:B)\leq I(A:B\tilde{B})=S_{A}-S_{\tilde{A}}+E_{W}(A:B)\leq 0$ (this phenomena is a generalization of a property of entanglement wedge: $A\cap B=\emptyset\Rightarrow\mathcal{E}_{A}\cap\mathcal{E}_{B}=\emptyset$ \cite{EW1}).
So we don't need to care about it.}
divided by $4G_{N}$. Then it immediately follows that $I(\tilde{A}:\tilde{B})=S_{\tilde{A}}+S_{\tilde{B}}-S_{\tilde{A}\tilde{B}}=0$.
On the other hand, $I(A:\tilde{A})=S_{A}+S_{\tilde{A}}-S_{A\tilde{A}}$
will be UV divergent because  $S_{A}$ and $S_{\tilde{A}}$ are itself divergent.
Note that the entanglement wedge cross section $E_{W}(A:B)=S_{A\tilde{A}}$
is always finite (assuming $A\cap B$ is empty). Thus subtracting
this term does not make $I(A:\tilde{A})$ finite. With the simple
setup described above, it can be written explicitly by
\begin{equation}
I(A:\tilde{A})=\frac{c}{3}\log\left[\frac{ld}{\epsilon^{2}}\right],
\end{equation}
where $\epsilon$ is the UV cutoff. After the phase transition,
we get a constant $I(A:\tilde{A})=2S_{A}=\frac{2c}{3}\log[\frac{l}{\epsilon}]$.
We plot the $I(A:\tilde{A})$ after subtracting out $2S_{A}$ in Fig.~(\ref{fig:MIAtildeA}).
Finally, $I(A:\tilde{B})$ is finite in general as usual for the
two subsystems separated from each other. Especially in AdS$_{3}$/CFT$_{2}$, $I(A:\tilde{B})$ always vanishes because the conformal symmetry allows us to set the subsystems in a symmetric
way so that $S_{A}+S_{\tilde{B}}=S_{\tilde{A}}+S_{B}=S_{A\tilde{B}}$.

The holographic entanglement of purification also satisfies an inequality called the strong superadditivity \cite{TaUm}. This property is not satisfied by the entanglement of purification for generic quantum states. Therefore this property can be regarded as a special property for holographic states. We will discuss this later in section (\ref{VioMonoSSA}).

\section{Computing EoP in Free Scalar Field Theory}

Here we present a general strategy to calculate the EoP in the ground state of a $1+1$ dimensional free scalar field theory. We discretize the field theory on a lattice and compute the EoP numerically. Our basic assumption is that since ground state wave functionals of free field theories are Gaussian, the wave functionals which appear after the purifications are also Gaussian. We also choose the minimal size ansatz. Under this assumption, we can calculate the EoP from matrix computations as we will explain below.

\subsection{Free Scalar Field Theory and Discretization}

Consider a free massive scalar field theory in $1+1$ dimension defined by the standard
Hamiltonian:
\be
H_0=\frac{1}{2}\int dx \left[\pi^2+(\de_x\phi)^2+m^2\phi^2\right].
\ee
We consider its lattice regularization by identifying $x=an$, where $a$ is the lattice spacing and $n=1,2,\ddd,N$ describes the position of each site (see e.g.\cite{Sha,Shb,ShTa}).
We define the discretized scalar field and its momentum at $n$-th site: $\phi_n=\phi(na)$ and $\pi_n=a\cdot \pi(na)$, which satisfy the canonical quantization condition $[\phi_n,\pi_{n'}]=i\delta_{n,n'}$. We impose the periodic boundary condition $\phi_{n+N}=\phi_n$ and $\pi_{n+N}=\pi_n$.

\begin{figure}
  \centering
  \includegraphics[width=6cm]{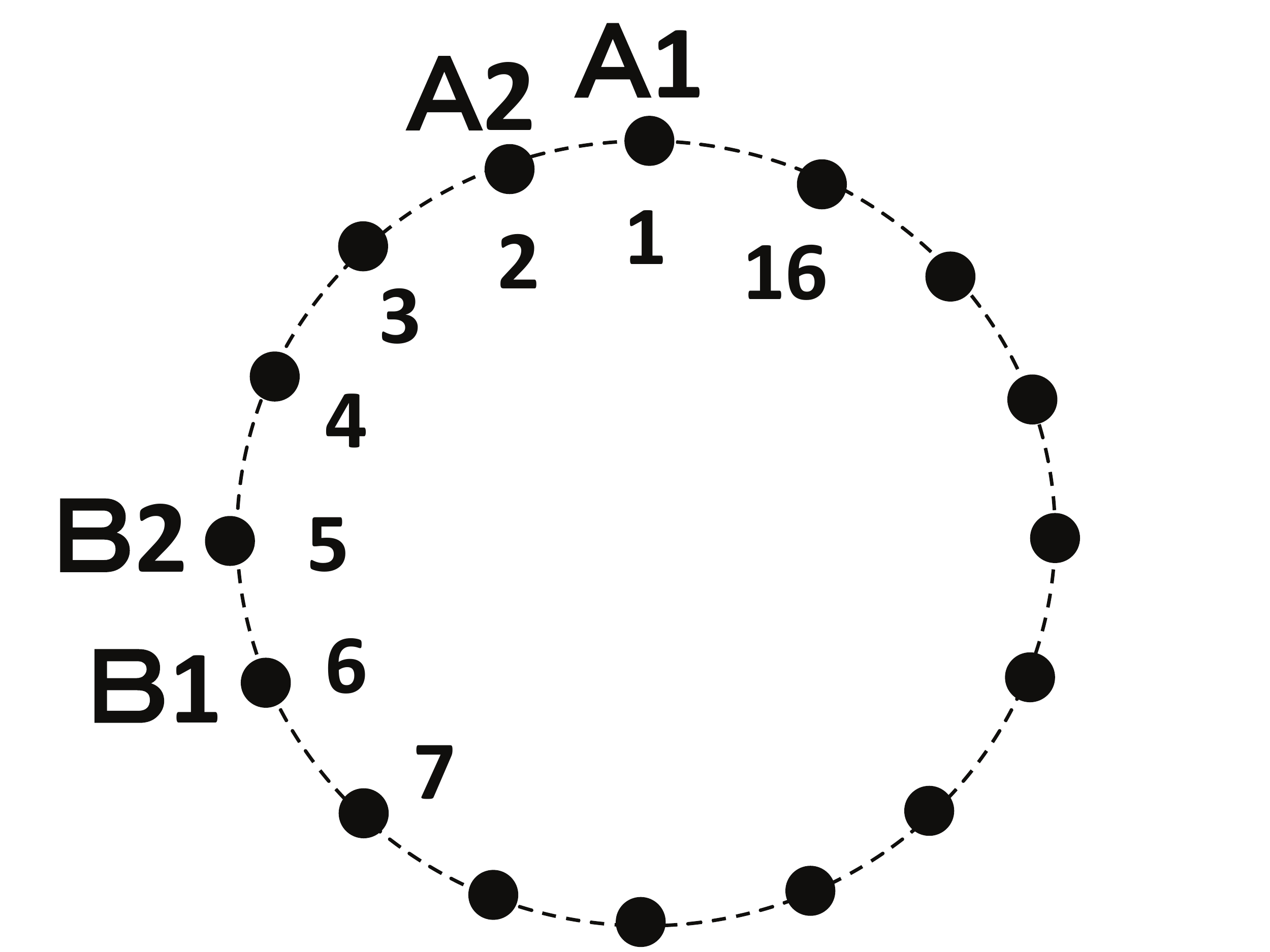}
  \caption{An example of the setup for our lattice model. We set $N=16$ and took $|A|=|B|=2$.
  The distance $d$ between $A$ and $B$ is $d=3$. The complement of $A$ and $B$, called $C$, consists of twelve lattice sites.}
\label{fig:setup}
  \end{figure}

Then the rescaled Hamiltonian $H=aH_0$ reads
\be
H=\sum^N_{n=1}\frac{1}{2}\pi_n^2+\sum^N_{n,n'=1}\frac{1}{2}\phi_n V_{nn'}\phi_{n'},
\ee
where the $N\times N$ matrix $V$ is given by
\be
V_{nn'}=N^{-1}\sum^{N}_{k=1} \left[a^2m^2+2\left(1-\cos\left(2\pi k/N\right)\right)\right]e^{2\pi ik(n-n')/N}.
\ee
The ground state wave function $\Psi_0$ of this lattice scalar theory is computed as
\be
\Psi_0[\phi]={\cal N}_0\cdot e^{-\frac{1}{2}\sum^N_{n,n'=1}\phi_n W_{mn}\phi_n'},
\ee
where the matrix $W$ is given by $\sqrt{V}$ or more explicitly:
\be
W_{nn'}=\frac{1}{N}\sum_{k=1}^N\s{a^2m^2+2\left(1-\cos\left(2\pi k/N\right)\right)}e^{2\pi ik(n-n')/N}.
\ee
Note that $W$ is a symmetric and real valued matrix. In the present paper, we will set
$a=1$ by rescaling the definition of the mass parameter $m$. In our actual numerical computations  we will always choose $N=60$ and consider five different masses $m=0.0001,0.001,0.01,0.1,1$. A sketch for $N=16$ can be found in Fig.(\ref{fig:setup}).

\subsection{Calculation of Entanglement Entropy}

We will follow the analysis in \cite{BKLS,Sha,Shb,ShTa} of computation of entanglement entropy in free scalar models. We decompose the Hilbert space ${\cal H}_{tot}$
as ${\cal H}_{tot}={\cal H}_A\otimes {\cal H}_B$ by choosing the subregion $A$ and its complement $B$ in a lattice system. The numbers of sites in $A$ and $B$ are called $|A|$ and $|B|$.

Consider a gaussian state $|\Psi\lb_{AB}$
in  ${\cal H}_{tot}$, which is in general written as follow:
\be
\Psi_{AB}={\cal N}_{AB}\cdot \exp\Biggl[-\frac{1}{2}(\phi_A\ \phi_B)
\left(\begin{array}{cc}
  A & B \\
  B^T & C
\end{array}\right)
\left(\begin{array}{c}
  \phi_A \\
  \phi_B
\end{array}\biggr)
\right].
\ee
We define the matrix $W$ and its inverse:
\ba
W=\left(\begin{array}{cc}
  A & B \\
  B^T & C
\end{array}\right),\ \ \ \
W^{-1}=
\left(\begin{array}{cc}
  D & E \\
  E^T & F
\end{array}\right),  \label{wdef}
\ea
where we have the obvious relations
\ba
AD+BE^T=B^T E+CF=1,\ \ \ AE+BF=B^TD+CE^T=0.
\ea
Note that for physically acceptable quantum states, the wave function should be
normalizable i.e. $W$ should be positive definite.

In this setup the entanglement entropy $S_A=S_B=-\mbox{Tr}[\rho_A\log\rho_A]$ is
computed as follows \cite{BKLS,Sha,Shb,ShTa}. First we compute the eigenvalues $\{\lambda_i\}$ of the matrix $\Lambda$ defined by
\be
\Lambda=-E\cdot B^T=D\cdot A-1, \label{eigen}
\ee
which is positive definite. The entanglement entropy is then computed by the formula
\be
S_A=S_B=\sum^{|A|}_{i=1}f(\lambda_i),  \label{xxx}
\ee
where
\be
f(x)=\log\frac{\s{x}}{2}+\s{1+x}\log\left(\frac{1}{\s{x}}+\frac{\s{1+x}}{\s{x}}\right).
\ee

\subsection{Calculations of EoP}

Now we are in a position to present how to calculate the EoP: $E_P(A:B)=E_P(\rho_{AB})$ defined by (\ref{EoP}) for the ground state $\Psi_0$ in our free scalar lattice model. We divide the total lattice system into subregions $A, B$ and $C$ such that ${\cal H}_{tot}={\cal H}_A\otimes {\cal H}_B\otimes {\cal H}_C$. We defined their lattice sizes to be $|A|,|B|$ and $|C|$.
In this setp, we would like to compute the EoP which measures a correlation between $A$ and $B$.

First, we write the ground state wave functionals in the following form:
\ba
\Psi_0[\phi_{AB},\phi_C]={\cal N}_0\cdot \exp\left[-\frac{1}{2}(\phi_{AB},\phi_C)
\left(\begin{array}{cc}
  P & Q \\
  Q^T & R
\end{array}\right)
\left(\begin{array}{c}
  \phi_{AB} \\
  \phi_C
\end{array}\right)
\right].\label{gab}
\ea
Note that the matrices $P,Q,R$ are all real valued; $P$ and $R$ are symmetric
matrices.

Then the reduced density matrix $\rho_{AB}=\mbox{Tr}_{\ti{A}\ti{B}}
\left[|\Psi_{A\ti{A}B\ti{B}}\lb \la\Psi_{A\ti{A}B\ti{B}}|\right]$  is obtained by integrating out $C$:
\ba
&& \rho_{AB}[\phi_{AB},\phi'_{AB}]\no
&&=\int D\phi_C \Psi^*_0[\phi_{AB},\phi_C]\cdot\Psi_0[\phi'_{AB},\phi_C]\no
&&\propto \exp\left[-\frac{1}{2}(\phi_{AB},\phi'_{AB})
\left(\begin{array}{cc}
  P-\frac{1}{2}QR^{-1}Q^T & -\frac{1}{2}QR^{-1}Q^T \\
  -\frac{1}{2}QR^{-1}Q^T & P-\frac{1}{2}QR^{-1}Q^T
\end{array}\right)
\left(\begin{array}{c}
  \phi_{AB} \\
  \phi'_{AB}
\end{array}\right)
\right]. \label{rab}
\ea

Our basic and crucial assumption is that
the optimal purified state $|\Psi_{A\ti{A}B\ti{B}}\lb$  in each setup, which minimizes $S_{A\ti{A}}$, is a gaussian state, described by the gaussian wave functional
\ba
&& \Psi_{A\ti{A}B\ti{B}}[\phi_{AB},\phi_{\ti{A}\ti{B}}] \no
&& ={\cal N}_{A\ti{A}B\ti{B}}\cdot \exp\left[
-\frac{1}{2}(\phi_{AB},\phi_{\ti{A}\ti{B}})
\left(\begin{array}{cc}
  J & K \\
  K^T & L
\end{array}\right)
\left(\begin{array}{c}
  \phi_{AB} \\
  \phi_{\ti{A}\ti{B}}
\end{array}\right)
\right],  \label{ababwf}
\ea
where $J$ and $L$ are real symmetric matrices and $K$ is a real matrix. For later use, we introduce the matrix $S$:
\be
S=
\left(\begin{array}{cc}
  J & K \\
  K^T & L
\end{array}\right).
\ee

Since the reduced density matrix $\rho_{AB}$ should agree with (\ref{rab}), we find the following two constraints:
\be
J=P,\ \ \ K L^{-1}K^T=QR^{-1}Q^T. \label{const}
\ee
With these constraints (\ref{const}) imposed, we can calculate the entanglement entropy $S_{A\ti{A}}=S_{B\ti{B}}$ from the total wave function $\Psi_{A\ti{A}B\ti{B}}$
(\ref{ababwf}) and minimize its value against the parameters in $K$ and $L$.
This is our basic strategy to calculate the EoP.

Here the gaussian ansatz of the purified state (\ref{ababwf}) is just an assumption which we cannot justify with any solid argument. However, it is natural to expect that the class of gaussian wave functionals are closed in themselves and that we may have only to take the minimization of $S_{A\ti{A}}$ within this class. Indeed as we will present below, this ansatz produces many reasonable results, being consistent with the general properties of EoP. Even if our expectation fails, our ``minimal gaussian EoP'' provides at least a useful upper bound of the actual EoP, which is defined by minimizations over all possible purifications.

\subsection{Symmetry Transformation}

In our computation of EoP, we can identify a symmetry transformation of the matrices $K$ and $L$
which do not change the value of $S_{A\ti{A}}$.

We take $P$ and $Q$ to be two non-degenerate matrices with the sizes $|\tilde{A}|$ and
$|\tilde{B}|$, respectively. We also introduce related matrices $\hat{P}$
(size $|A|+|\tilde{A}|$) and $\hat{Q}$ (size $|B|+|\tilde{B}|$) defined by
\be
\hat{P}=\left(\begin{array}{cc}
  I_{|A|} & 0  \\
  0 & P
\end{array}\right),\ \ \ \
\hat{Q}=\left(\begin{array}{cc}
  I_{|B|} & 0 \\
  0 & Q
\end{array}\right),
\ee
where $I_{|A|,|B|}$ are the identity matrices.

The symmetry transformation is given by
\ba
&& J\to J,\ \ \ \ K\to K
\left(\begin{array}{cc}
  P^T & 0 \\
  0 & Q^T
\end{array}\right),\ \ \
L\to \left(\begin{array}{cc}
  P & 0 \\
  0 & Q
\end{array}\right) L
\left(\begin{array}{cc}
  P^T & 0 \\
  0 & Q^T
\end{array}\right).  \label{symt}
\ea
To see if these transformations indeed do not change the entanglement entropy $S_{A\ti{A}}$,
we can look at the matrix $W$ obtained by rearranging $(J,K,L)$ as follows:
\be \label{W}
W=
\left(\begin{array}{cccc}
  J_{AA} & K_{A\ti{A}} & J_{AB} & K_{A\ti{B}}\\
  K_{\ti{A}A} & L_{\ti{A}\ti{A}} & K_{\ti{A}B} & L_{\ti{A}\ti{B}}\\
  J_{BA} & K_{B\ti{A}} & J_{BB} & K_{B\ti{B}}\\
  K_{\ti{B}A} & L_{\ti{B}\ti{A}} & K_{\ti{B}B} & L_{\ti{B}\ti{B}}.
\end{array}\right)
\equiv\left(\begin{array}{cc}
  A & B \\
  B^T & C
\end{array}\right),
\ee
where we decompose $(J,K,L)$ based on the indices $A$, $\ti{A}$, $B$ and $\ti{B}$ in an
obvious way. The sizes of the matrices $A$, $B$ and $C$ are $(|A|+|\ti{A}|)\times (|A|+|\ti{A}|)$,
$(|A|+|\ti{A}|)\times (|B|+|\ti{B}|)$, $(|B|+|\ti{B}|)\times (|B|+|\ti{B}|)$, respectively.

In terms of $(A,B,C)$, the transformations are expressed as
\ba
&& A\to \hat{P}A\hat{P}^T,\ \ \ B\to \hat{P}B\hat{Q}^T,\ \ \ C\to \hat{Q}C\hat{Q}^T,\no
&& D\to (\hat{P}^T)^{-1}D\hat{P}^{-1},\ \ \ E\to (\hat{P}^T)^{-1}E\hat{Q}^{-1},\ \ \ F\to (\hat{Q}^T)^{-1}F\hat{Q}^{-1}.\no
\ea
 Thus $\Lambda=-E\cdot B^T$ is mapped by the similarity transformation $\Lambda\to (P^T)^{-1}\Lambda P^T$ and thus $S_{A\ti{A}}$, computed from the formula (\ref{xxx}), does not change.

By using this symmetry, we can reduce the number of parameters in $K$ and $L$ which we need to minimize to $|\ti{A}|^2+|\ti{B}|^2$.

\subsection{Minimal Gaussian Ansatz}\label{MinAnsatz}

Even if we assume the Gaussian ansatz, still it looks hopeless to numerically calculate the EoP because the sizes of matrices $K$ and $L$ can be infinite. Therefore we adopt a finite size ansatz, especially the minimal size one given by $|\ti{A}|=|A|$ and $|\ti{B}|=|B|$.
We call this the minimal Gaussian ansatz. This minimal ansatz is employed to produce our numerical results of EoP, which will be presented in coming sections.

Even though we do not have a full justification of this ansatz, we have numerical supporting evidence that this ansatz can give an exact answer: even if we start with larger sizes of the purification spaces $|\ti{A}|> |A|$ and $|\ti{B}|> |B|$, we will get back to the minimal one $|\ti{A}|=|A|$ and $|\ti{B}|=|B|$ after the minimization, as we will present in the section~(\ref{reduction}).

In this minimal ansatz, we can reduce the matrix $K$ into the following form by taking advantage of the symmetry transformation (\ref{symt}):
\ba
K=
\left(\begin{array}{cc}
  I_{|A|} &  K_{A\ti{B}} \\
  K_{B\ti{A}}   & I_{|B|}
\end{array}\right), \label{minima}
\ea
which has $2|A||B|$ parameters.  The matrix $L$ is completely determined from $K$ by
the constraint (\ref{const}). Thus the numerical computation of EoP in our setup requires
the minimization of $S_{A\ti{A}}$ over the $2|A||B|$ parameters.

In our explicit numerical analysis
presented below we will focus on the cases $(|A|,|B|)=(1,1), (1,2)$ and $(2,2)$ with the total number of lattice sites $N=60$.

\section{Numerical Results of EoP}\label{NREoP}

Now we are prepared to present our numerical results of EoP in our free scalar theory.
We choose the total lattice size to be $N=60$ and the subsystem sizes to be $(|A|,|B|)=(1,1),(1,2),(2,2)$.
We perform the numerical computation of EoP $E_P(A:B)$ for five different scalar field masses $m=0.0001,0.001,0.01,0.1,1$  (we set $a=1$). We are interested in how $E_P(A:B)$
depends on the distance $d$ between $A$ and $B$ (refer to Fig.~(\ref{fig:setup})).
We employ the minimal Gaussian ansatz (\ref{minima}). Thus we have only to minimize $S_{A\ti{A}}$ with respect to the $2|A||B|$ parameters in $K_{A\ti{B}}$ and $K_{B\ti{A}}$ as the matrix $L$ is completely determined by $K$. In the final subsection, we will present some evidence that supports the minimal ansatz.

\subsection{Example $1$: $|A|=|B|=1$}\label{argument}

Let us start with the smallest subsystems $|A|=|B|=1$. In this case, we need to minimize
with respect to two real parameters $K_{A\ti{B}}=x_1$ and $K_{B\ti{A}}=x_2$. In our explicit numerical calculations, we always find $x_1=x_2$ at any minimum points. This can be understood from the obvious Z$_2$ symmetry in the original system which replaces $A$ with $B$ and vice-versa. This symmetry leads to the symmetry which exchanges $(A,\ti{A})\lr (B,\ti{B})$ in the purified system.

\begin{figure}[ht]
  \centering
  \includegraphics[width=5cm]{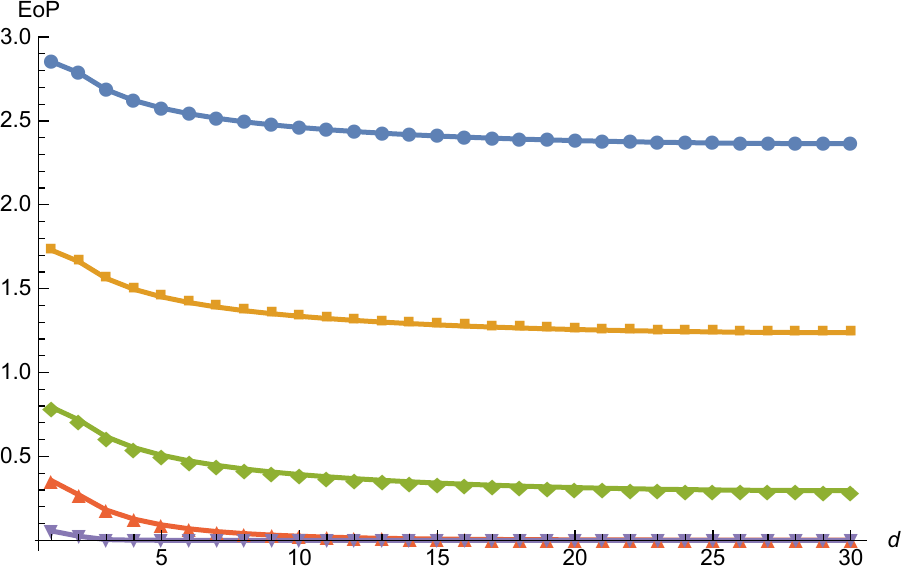} \hspace{1cm}
  \includegraphics[width=6.0cm]{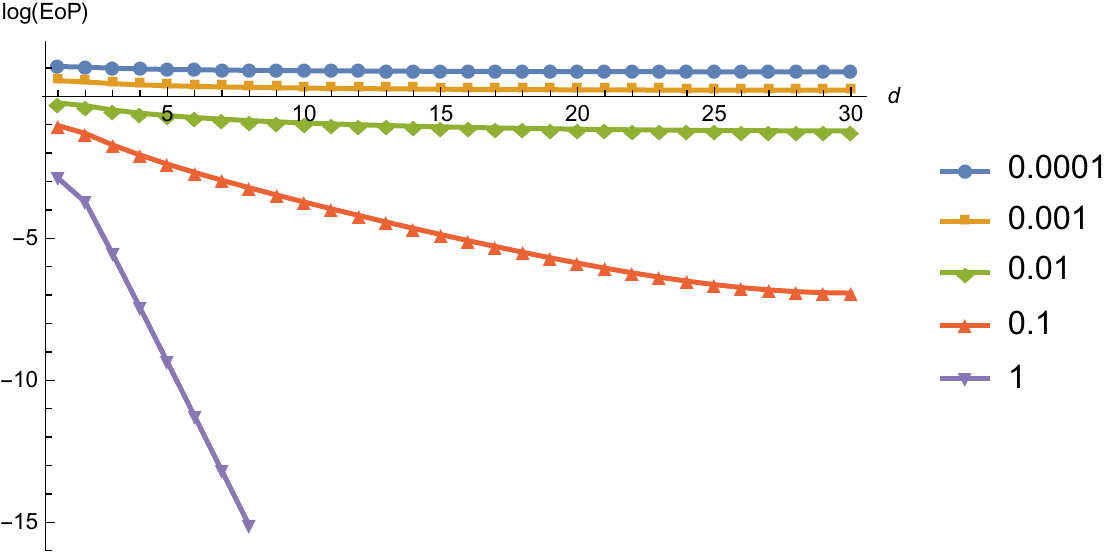}\\ \vspace{5mm}
  \includegraphics[width=5cm]{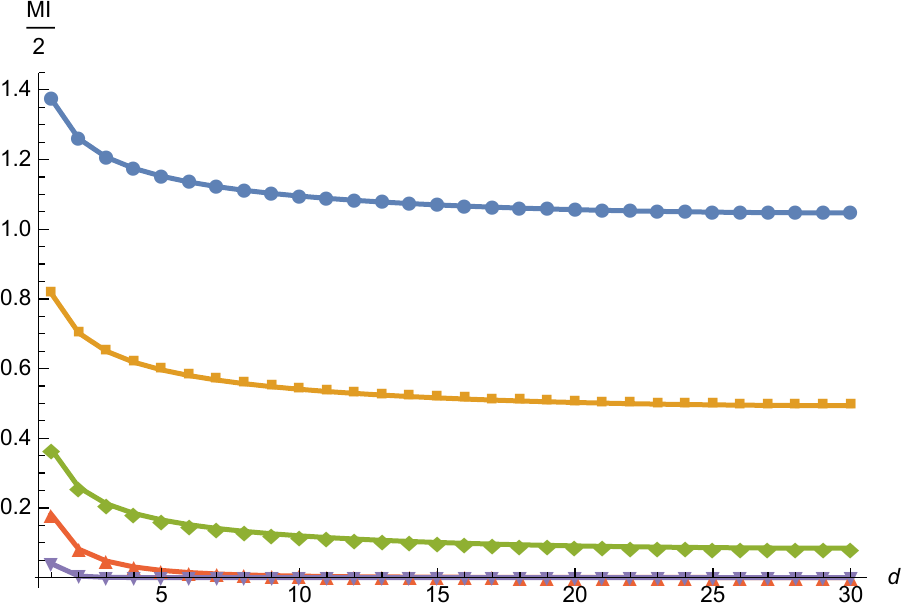}\hspace{1cm}
  \includegraphics[width=6.0cm]{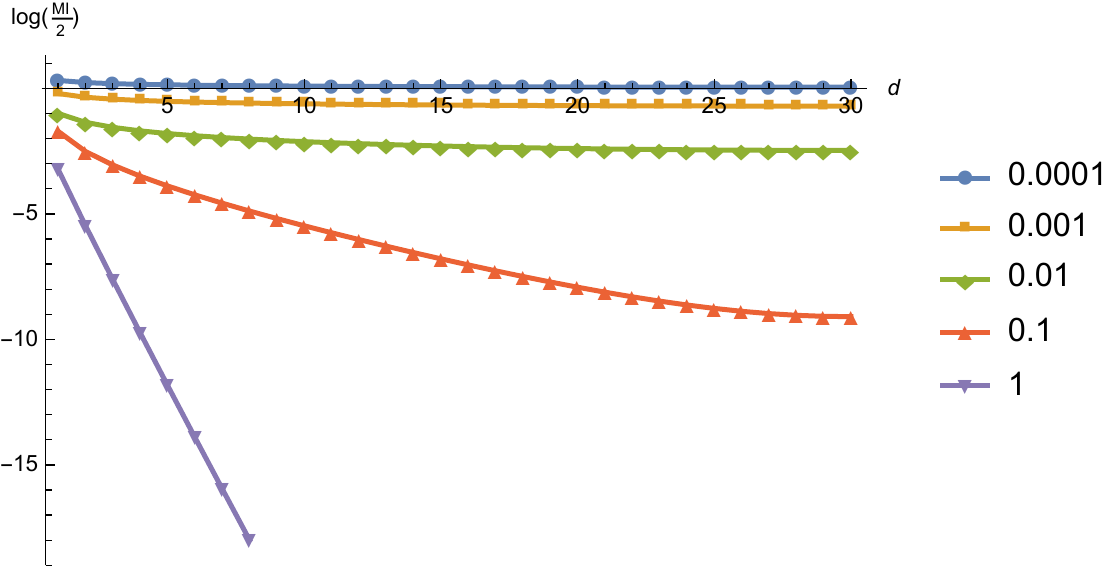}\\
  \caption{The plots of EoP (upper-left graphs) and a half of mutual information $I(A:B)$
    (lower-left graphs) in the setup of $|A|=|B|=1$ as a function of $d$, which is the distance between $A$ and $B$ (we took $1\leq d \leq 30$). The right ones are obtained by taking the logarithms of the left ones. In each graph, from the above to the bottom, the mass varies $m=0.0001,0.001,0.01,0.1,1$.}
\label{fig:A1B1}
\end{figure}

Our numerical results of EoP and a half of the mutual information are plotted in Fig.~(\ref{fig:A1B1}). Note that the former should always be larger than the latter as in 
(\ref{NIuDA}) and this is
indeed true in our numerical results. As is clear from the graphs, both of EoP and mutual information are monotonically decreasing as the distance $N_d$ gets larger. As the mass gets larger, both graphs change from the power law decay to the exponential decay, as naturally expected.

\begin{figure}[ht]
  \centering
  \includegraphics[width=6cm]{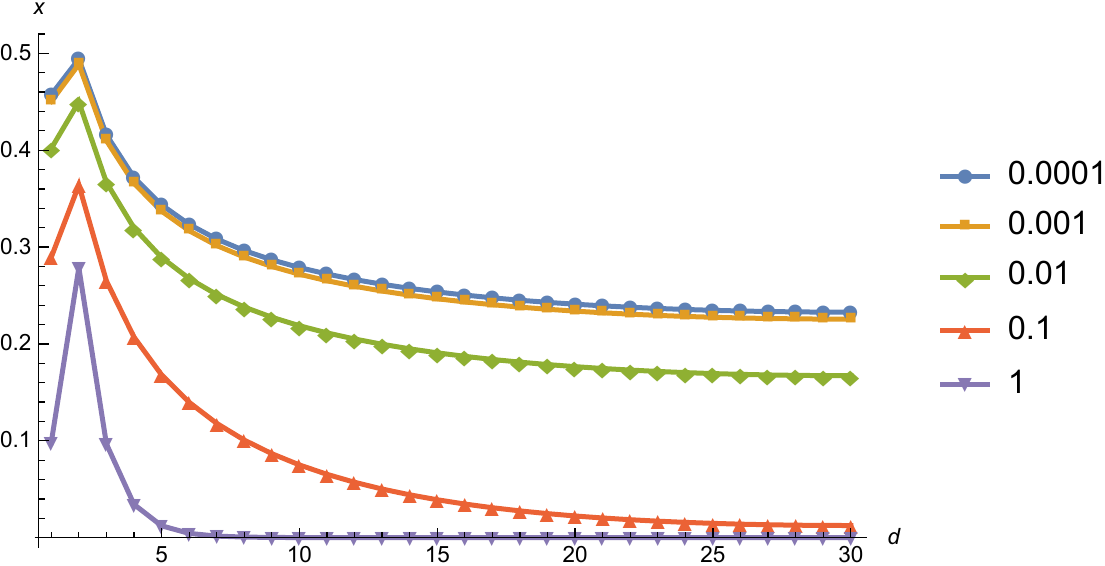}
    \caption{The plots of the optimized values of the parameter $x$ which give the minimum of
    $S_{A\tilde{A}}$ in the setup of $|A|=|B|=1$. In this graph, from the above to the bottom, the mass varies $m=0.0001,0.001,0.01,0.1,1$.}
\label{fig:A1B1X}
  \end{figure}

We also plotted the values of $x_1=x_2$ which minimize  $S_{A\ti{A}}$
in Fig.~(\ref{fig:A1B1X}). It is intriguing to notice that each graph has always a peak at
$d=2$. We can explain this behavior as follows. When $d=2$, $A$ and $B$ are the next to the nearest neighbor and there is a single lattice site, called $C$, between $A$ and $B$. It is clear that in the original wave functional, the entanglement between $A$ and $C$ and that between $B$ and $C$ are both equally very strong. Therefore in the purified state $|\Psi\lb_{A\ti{A}B\ti{B}}$, we can expect that both $\ti{A}$ and $\ti{B}$ are closely
related to the site $C$. This means that the correlation between $A$ and $\ti{B}$ and the one
between $B$ and $\ti{A}$ get enhanced for $d=2$. On the other hand, when $d=1$ and $d\geq 3$, a similar consideration does not lead to any clear enhancement. Indeed the parameters $K_{A\ti{B}}=x_1$ and $K_{B\ti{A}}=x_2$ are obviously responsible for these correlations. This explains the peak at $d=2$.

\subsection{Example $2$: $|A|=1,\ |B|=2$}

Next we proceed to the case $|A|=1,\ |B|=2$. The two sites in $B$ are separately called $B_1$ and $B_2$. In this case we minimize $S_{A\ti{A}}$ with respect to the four parameters $(x,y,z,w)$ in the matrix $K$ of the form (the indices of $K$ are arranged in the order $AB_1B_2 \times \ti{A}\ti{B}_1\ti{B}_2$) \footnote{ Here we have chosen a difference ansatz than (\ref{minima}) for our convenience. Both should
give the same results for EoP.}
\be
K=
\left(\begin{array}{ccc}
 1 & 0 & x \\
 y & 1 & 0 \\
 z & w & 1
\end{array}\right).
\ee

\begin{figure}[ht]
  \centering
  \includegraphics[width=5cm]{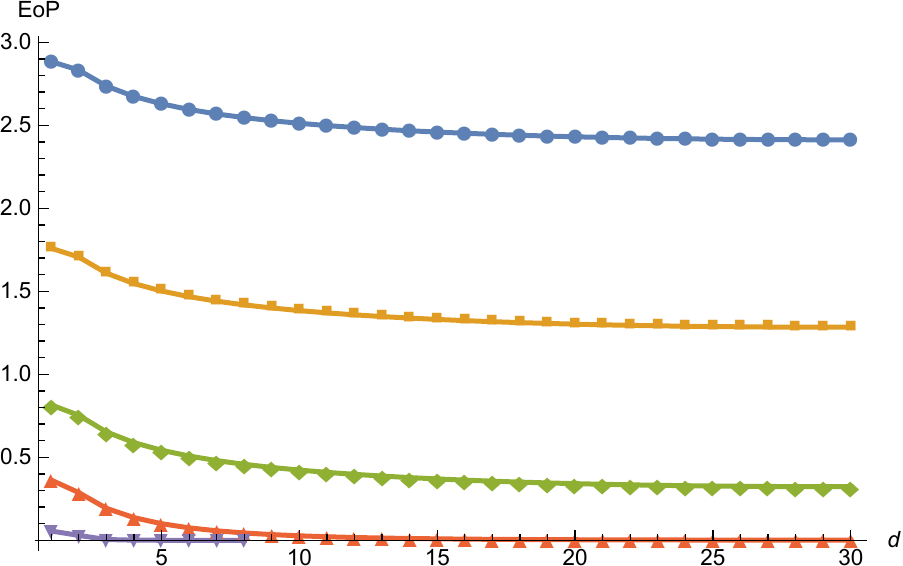}
  \includegraphics[width=6.0cm]{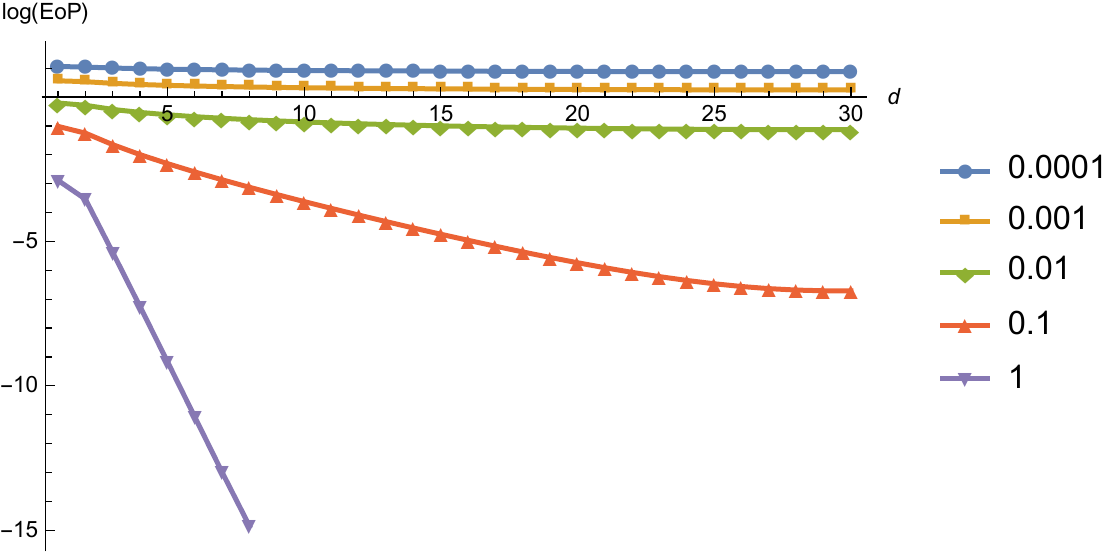}\\
  \includegraphics[width=5cm]{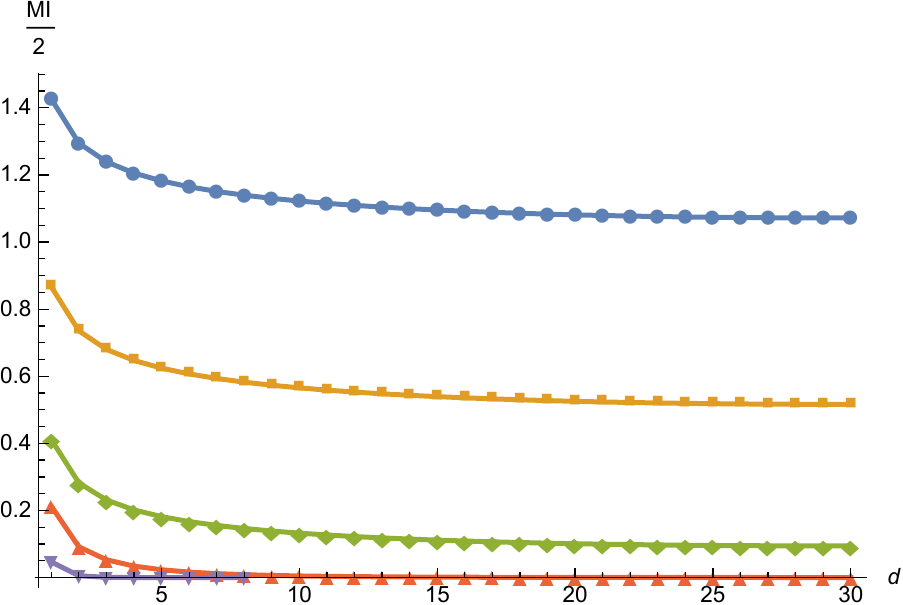}
  \includegraphics[width=6.0cm]{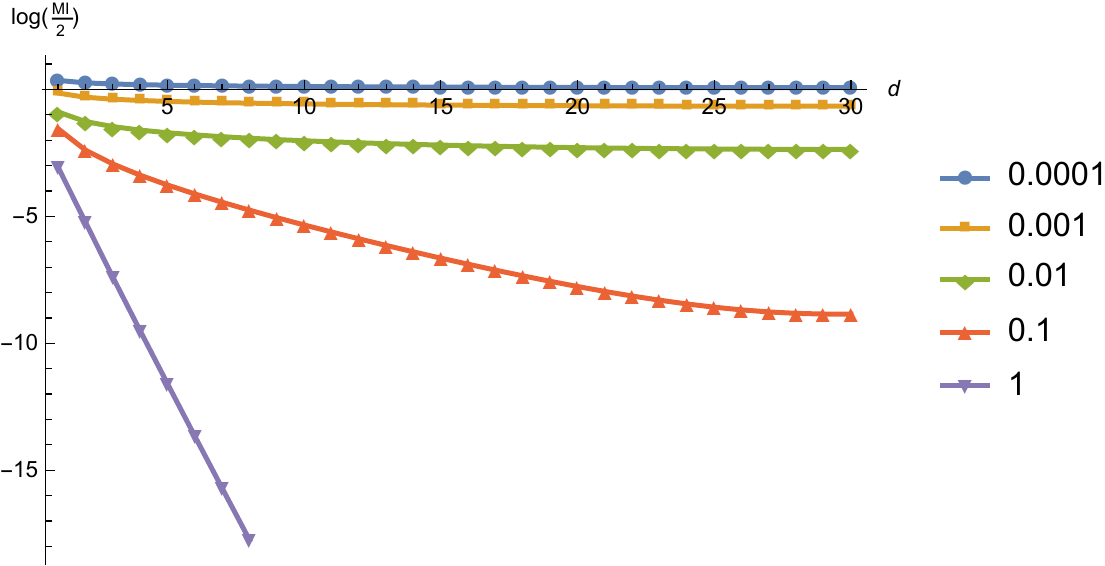}\\

    \caption{The plots of EoP (upper graphs) and a half of mutual information $I(A:B)$
    (lower graphs) as a function of $d$ in the setup of $|A|=1,\ |B|=2$. The right ones are obtained by taking their logarithms. In each graph, from the above to the bottom, the mass varies $m=0.0001,0.001,0.01,0.1,1$.}
\label{fig:A1B2}
  \end{figure}

The results of EoP and a half of mutual information are plotted in Fig.~(\ref{fig:A1B2}). Qualitative behaviors are very similar to the previous ones
for $|A|=|B|=1$. As follows from the extensivity of EoP (\ref{NIuDA}), the result for $|A|=1,\ |B|=2$ is larger than that for $|A|=|B|=1$ with the same mass $m$ and $d$.

\begin{figure}[ht]
  \centering
  \includegraphics[width=4cm]{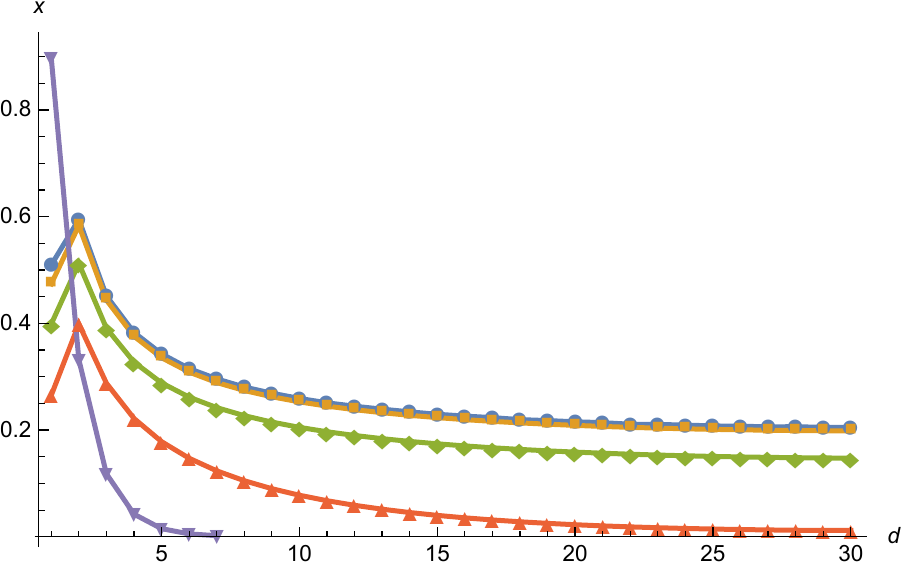}
  \includegraphics[width=4cm]{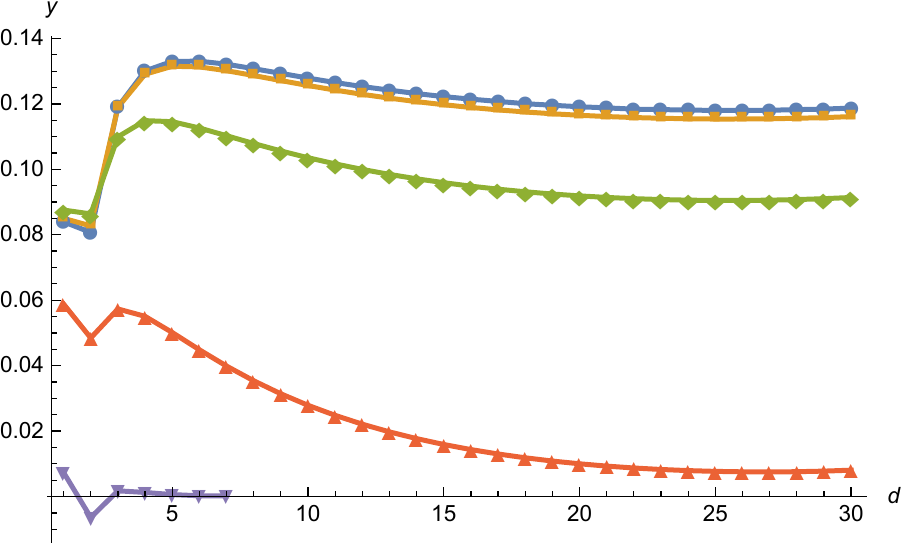}
  \includegraphics[width=4cm]{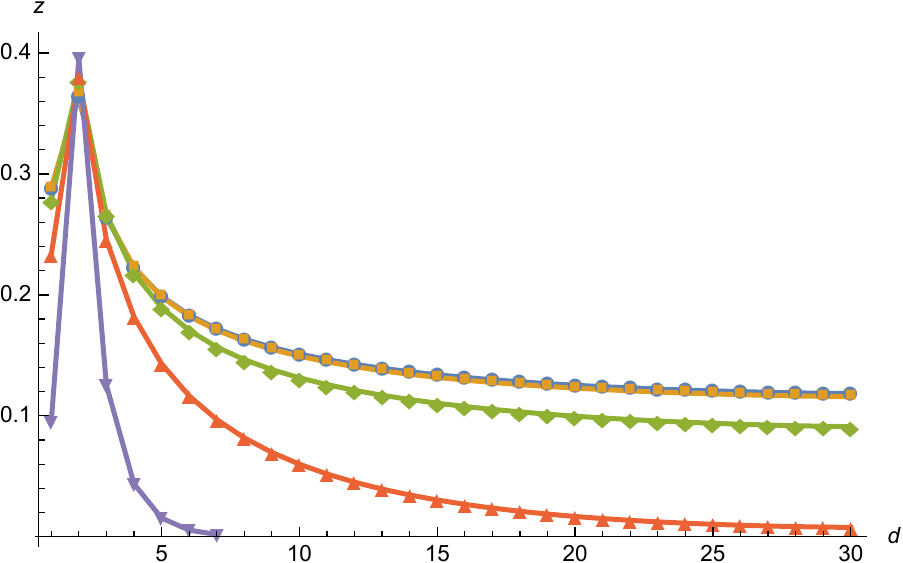}
  \includegraphics[width=4cm]{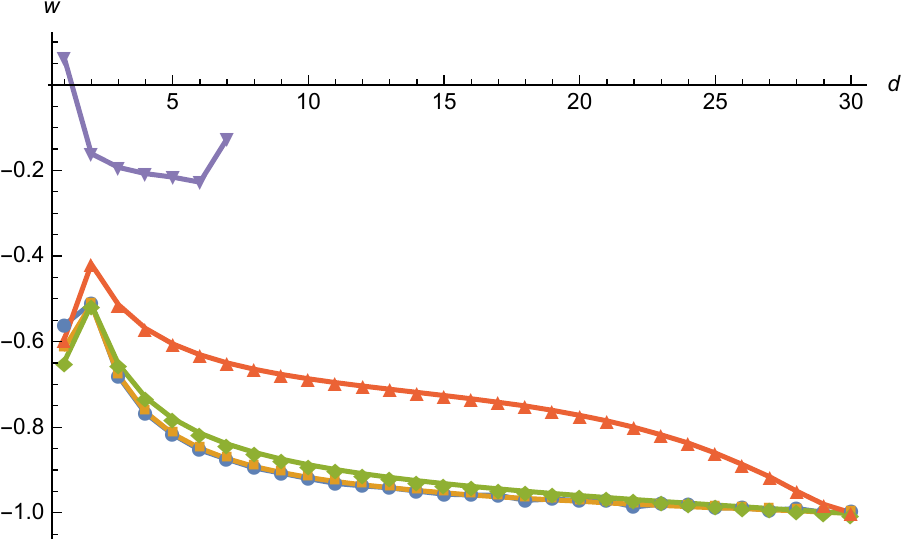}\\
    \caption{The plots of the optimized value of the parameter $x,y,z,w$ which gives the minimum of  $S_{A\tilde{A}}$ in the setup of $|A|=1,\ \ |B|=2$. In the first three graphs, from the above to the bottom, the mass varies $m=0.0001,0.001,0.01,0.1,1$. In the final graph the order is opposite.}
\label{fig:A1B2X}
  \end{figure}

We also plotted the values of $(x,y,z,w)$ where $S_{A\ti{A}}$ gets minimized in Fig.~(\ref{fig:A1B2X}). As the graphs show, the behaviors of $x$ and $z$ are similar to that of $x$ in the previous case $|A|=|B|=1$. Since $x$ (and $z$) here are related to the correlation between $A$ and $\ti{B}_2$ (and $B_2$ and $\ti{A}$), this is enhanced because there is only a single site between $A$ and $B_2$ (refer to Fig.~(\ref{fig:setup})) in the same way as before.
On the other hand, the values $y$ and $w$ are related to the correlations which are rather
suppressed by this effect.

\subsection{Example $3$: $|A|=2,\ |B|=2$}

Next we proceed to the case $|A|=|B|=2$. The two sites in $A$ and $B$ are separately called $A_1,A_2$ and $B_1,B_2$ (refer to Fig.~(\ref{fig:setup}) again). Note that, constrained by the resources at our disposal, this is the largest size of subsystems we consider in this paper for our convenience. We expect this example can have some features of field theory limits more the than other examples already discussed.

In this case we minimize $S_{A\ti{A}}$ with respect to the matrix $K$ of the form (the indices of $K$ are arranged in the order $A_1A_2B_1B_2 \times \ti{A}_1\ti{A}_2\ti{B}_1\ti{B}_2$) given by:
\be
K=
\left(\begin{array}{cccc}
 1 & 0 & x & y\\
 0 & 1 & z & w\\
 x' & y' & 1 & 0\\
 z' & w' & 0 & 1
\end{array}\right).
\ee
As we can also confirm numerically, the symmetry which exchanges $A$ and $B$ allows us to set $K_{A\ti{B}}=K_{B\ti{A}}$ , or equally $x=x', y=y', z=z'$ and $w=w'$. Thus to calculate the EoP, we need to minimize $S_{A\ti{A}}$ against four parameters $(x,y,z,w)$.

\begin{figure}[h!]
  \centering
  \includegraphics[width=5cm]{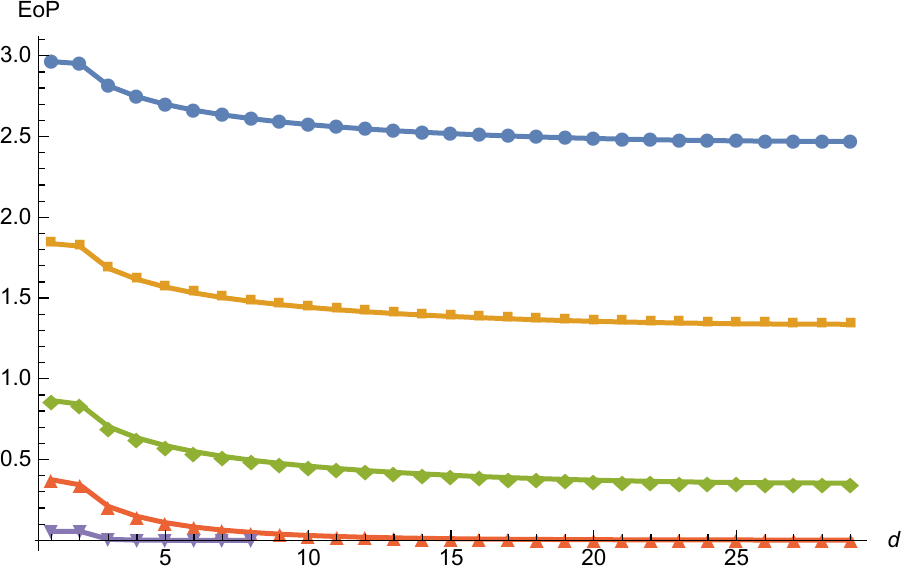}
  \includegraphics[width=6.0cm]{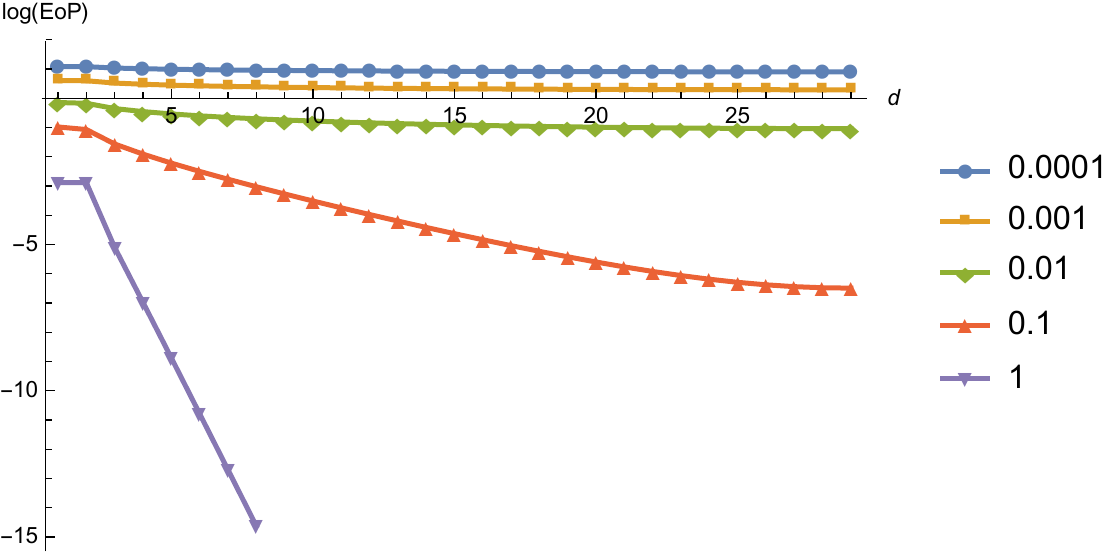}\\
  \includegraphics[width=5cm]{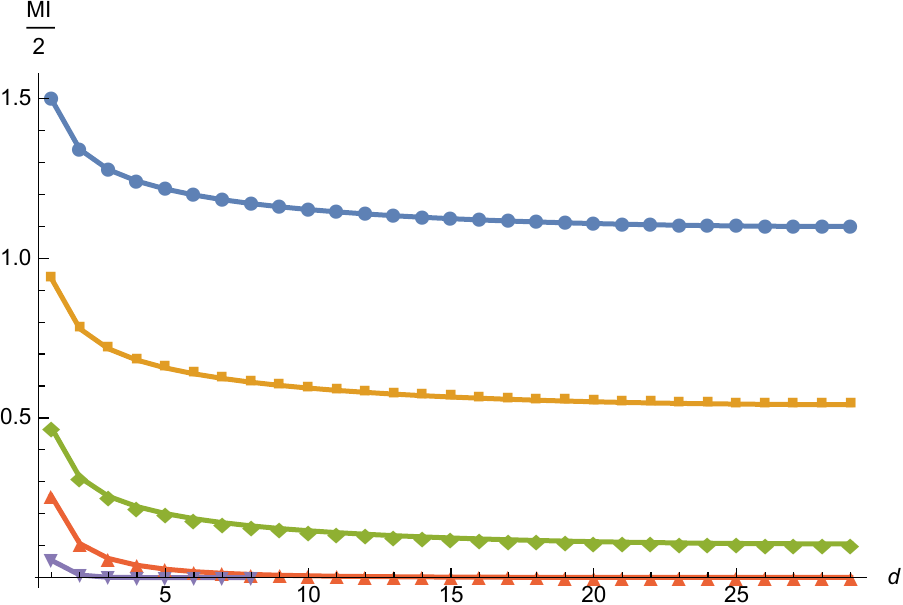}
  \includegraphics[width=6.0cm]{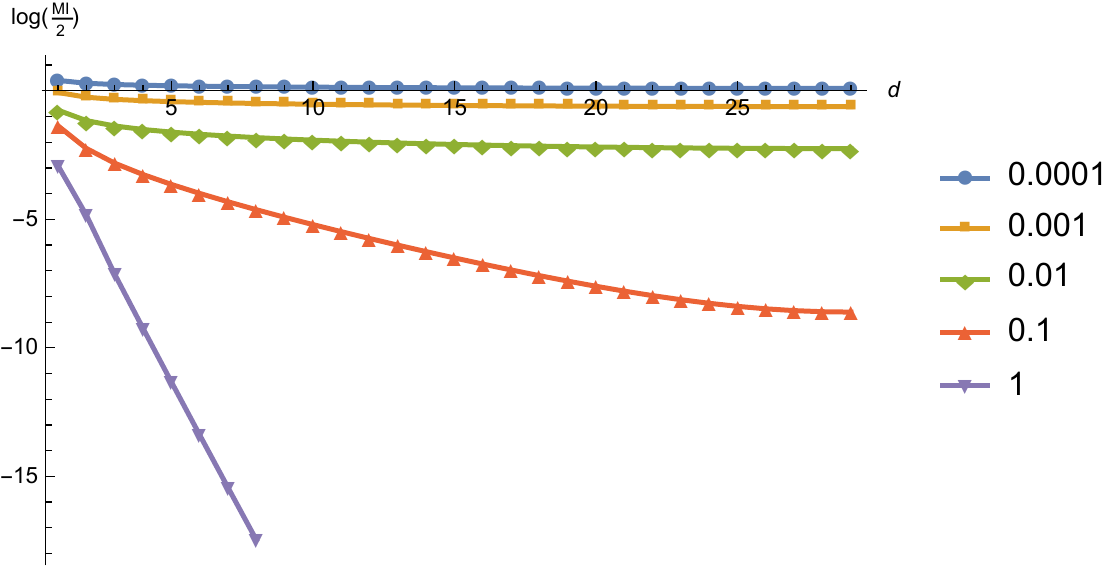}\\

    \caption{The plots of EoP (upper graphs) and a half of mutual information $I(A:B)$
    (lower graphs) as a function of $d$ in the setup of $|A|=|B|=2$. The right ones are obtained by taking their logarithms. In each graph, from the above to the bottom, the mass varies $m=0.0001,0.001,0.01,0.1,1$.}
\label{fig:A2B2}
  \end{figure}

After this minimization, we obtain the results of EoP in Fig.~(\ref{fig:A2B2}). By comparing them with previous ones, we can confirm the extensivity of EoP (\ref{NIuDA}).
Also both the EoP and mutual information are again monotonically decreasing. As the mass increases, the power law decay gets changed into an exponential decay.
However, we now notice an important difference between the EoP and the mutual information:
the values of EoP at $d=1$ and $d=2$ are almost the same, while those of
the mutual information are different. This plateaux in the EOP qualitatively looks similar to
what we observe in the holographic EoP, where there occurs a phase transition (refer to
Fig.~(\ref{fig:PhaseTransition})). Indeed the phase transition point is $d_*=(\s{2}-1)l$, where $l$ is the size of the subsystem $A$ and $B$. In our example, we took $l=2$ and thus $d_c\sim 1$, which is consistent with the above behavior.

\begin{figure}[ht]
  \centering
  \includegraphics[width=4cm]{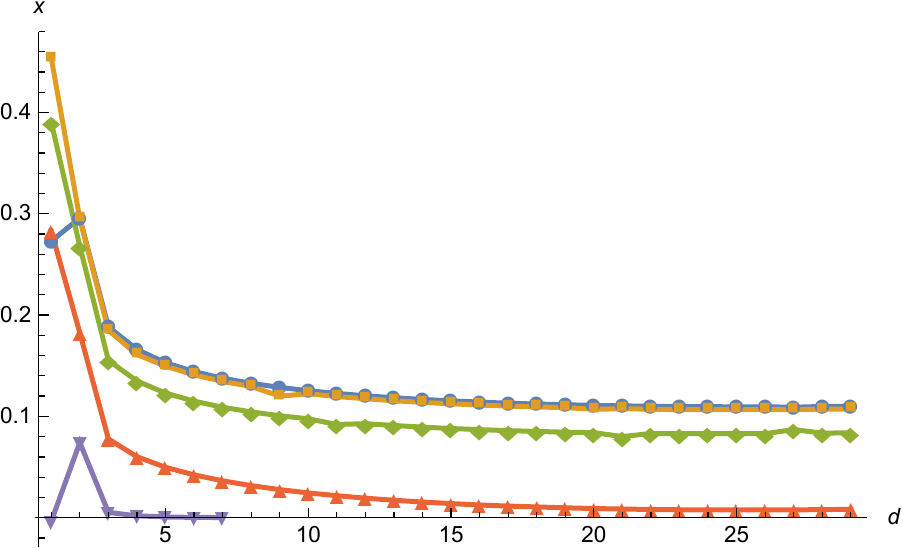}
  \includegraphics[width=4cm]{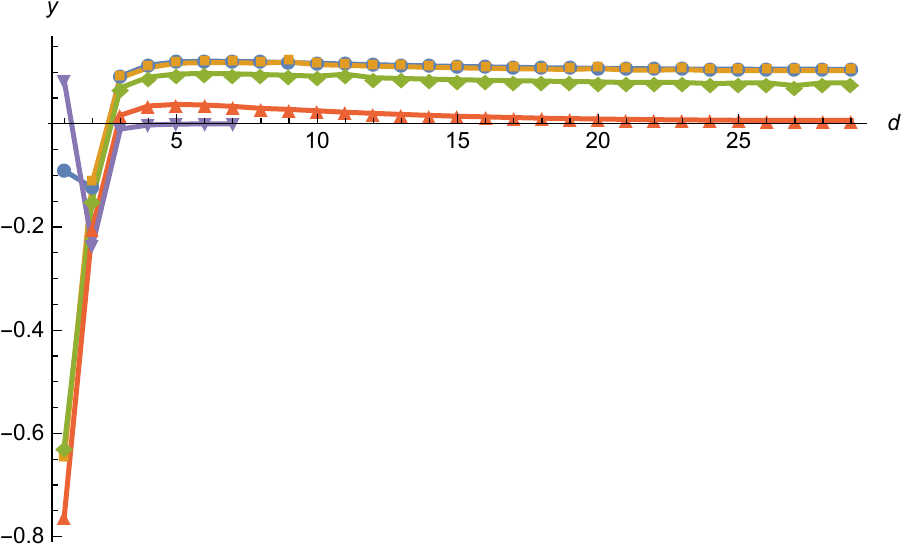}
  \includegraphics[width=4cm]{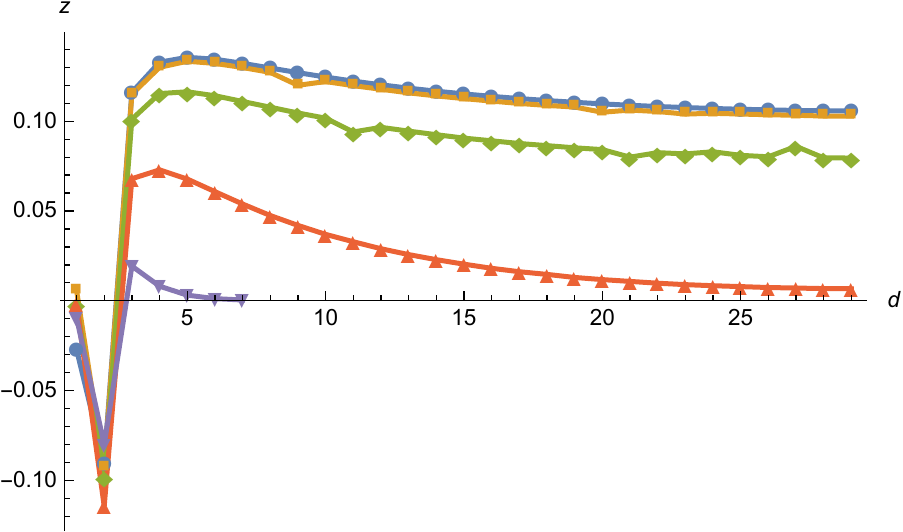}
  \includegraphics[width=4cm]{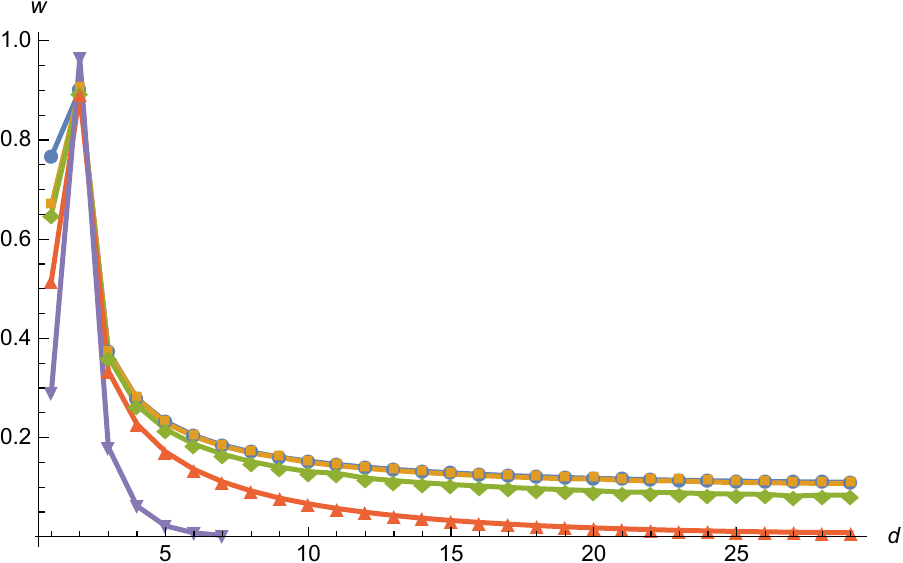}\\
    \caption{The plots of the optimized value of the parameter $x,y,z,w$ which gives the minimum of  $S_{A\tilde{A}}$ in the setup of $|A|=|B|=2$. In the first graph, from the above to the bottom, the mass varies $m=0.0001,0.001,0.01,0.1,1$.}
\label{fig:A2B2X}
  \end{figure}

It is also intriguing to examine the behavior of parameters $(x,y,z,w)$ at the minimum points.
They are plotted in Fig.~(\ref{fig:A2B2X}). First of all, we note that $w$ has a clear peak at
$d=2$ as in Fig.~(\ref{fig:A1B1X}). The reason for this peak is the same as
that of $x$ in $|A|=|B|=1$ case: $A_2$ get strongly correlated with $\ti{B}_2$ through the vacant site.  This effect highly reduces the correlation between $A_2$ and $\ti{B}_1$ and thus
the absolute values of $z$ behave in an opposite way. The behavior of $x$ and $y$ are roughly in the middle between these two. We also find from our numerical data that as $d$ gets larger,
the four parameters get closer to each other $x\simeq y\simeq z\simeq w$. This can be easily understood because when $A$ and $B$ are most separated, all four possible correlations
$A_1-B_1,\ A_1-B_2,\ A_2-B_1$ and $A_2-B_2$ should be strong in the same magnitude.

\subsection{Numerical Evidences for Minimal Ansatz} \label{reduction}

In this section, we would like to present numerical evidence that the minimal ansatz (\ref{minima}) discussed in sections~(\ref{MinAnsatz}), which was  employed for all our numerical computations, is sufficient to produce the correct EoP. We discuss this in details for $|A|=|B|=1$ case first. For the other two cases, the arguments will follow analogously.

 For this case $|\tilde A|=|\tilde B|=1$ is the minimal ansatz.  Now we try to increase the dimensions of this auxiliary Hilbert space. We  consider $|\tilde A|=|\tilde B|=2. $  The matrix $W$ in (\ref{W})  takes the following form ( elements are in the order $A\tilde A B\tilde B$),
\begin{align}
\begin{split}\label{largeansatz}
W=\left(
\begin{array}{cccccc}
 J_{AA} & K_{A\tilde A_{1}} & K_{A\tilde A_{2}} & J_{AB} & K_{A\tilde B_{1}} & K_{A\tilde B_2} \\
 K_{\tilde A_{1} A} & L_{\tilde A_1\tilde A_1} & L_{\tilde A_1\tilde A_2} & K_{\tilde A_{1} B} & L_{\tilde A_1\tilde B_{1}} & L_{\tilde A_{1}\tilde B_{2}} \\
 K_{\tilde A_2 A} & L_{\tilde A_2\tilde A_1} & L_{\tilde A_2\tilde A_2} & K_{\tilde A_2 B} & L_{\tilde A_2\tilde B_{1}} & L_{\tilde A_2\tilde B_{2}} \\
 J_{BA} & K_{B\tilde A_{1}} & K_{B\tilde A_{2}} & J_{BB} & K_{B\tilde B_{1}} & K_{B\tilde B_{2}} \\
 K_{\tilde B_1A} & L_{\tilde B_1\tilde A_{1}} & L_{\tilde B_{1}\tilde A_{2}} & K_{\tilde B_1B} & L_{\tilde B_{1}\tilde B_{1}} & L_{\tilde B_1\tilde B_2} \\
 K_{\tilde B_2A} & L_{\tilde B_2\tilde A_{1}} & L_{\tilde B_2\tilde A_{2}} & K_{\tilde B_2B} & L_{\tilde B_2\tilde B_{1}} & L_{\tilde B_2\tilde B_{2}} \\
\end{array}
\right).
\end{split}
\end{align}
We can easily see that the minimal ansatz is contained in this.  We set some of the entries to zero such that the $W$ matrix takes the following form,
\begin{align}
\begin{split}
W=\left(
\begin{array}{cccccc}
 J_{AA} & K_{A\tilde A_{1}} & 0& J_{AB} & K_{A\tilde B_{1}} & 0 \\
 K_{\tilde A_{1} A} & L_{\tilde A_1\tilde A_1} & 0& K_{\tilde A_{1} B} & L_{\tilde A_1\tilde B_{1}} & 0 \\
 0 & 0 & L_{\tilde A_2\tilde A_2} & 0 &0 & L_{\tilde A_2\tilde B_{2}} \\
 J_{BA} & K_{B\tilde A_{1}} & 0 & J_{BB} & K_{B\tilde B_{1}} & 0 \\
 K_{\tilde B_1A} & L_{\tilde B_1\tilde A_{1}} & 0 & K_{\tilde B_1B} & L_{\tilde B_{1}\tilde B_{1}} & 0 \\
 0& 0& L_{\tilde B_2\tilde A_{2}} & 0 & 0 & L_{\tilde B_2\tilde B_{2}} \\
\end{array}
\right).
\end{split}
\end{align}
From this it is evident that, $\tilde A_2 $ and $\tilde B_{2}$ do
not remain entangled with the composite $A, B,\tilde A_1, \tilde B_1$  system.  One can  then recover the previous results for EoP by setting $K_{A,\tilde A_1}=K_{B,\tilde B_1}=1$ and $K_{A, \tilde B_1}=K_{B,\tilde A_1}=x, L_{\tilde A_2,\tilde B_2}=0$ and minimizing over the parameter $x$ regardless of the values of $L_{\tilde A_2, \tilde A_2}, L_{\tilde A_{2}\tilde B_2}.$ ( $L_{\tilde A_2 \tilde B_2}=0$ makes $\tilde A_2$ independent of $\tilde B_2$ hence the $S_{A\tilde A_1\tilde A_2}$ naturally coincides with $ S_{A\tilde A_1}.$)   Now we want to  check that even if we start from (\ref{largeansatz}),  minimization of  $S_{A\tilde A_1\tilde A_2}$ will demand that   $\tilde A_2, \tilde B_2$ should decouple from the $A,B,\tilde A_1, \tilde B_1.$ To check this numerically we adopt the following strategy. For our case the $K$ matrix is the following,
\begin{align}
\begin{split}
K=\left(
\begin{array}{cccccc}
 K_{A\tilde A_{1}} & K_{A\tilde A_{2}} & K_{A\tilde B_{1}} & K_{A\tilde B_2} \\
 K_{B\tilde A_{1}} & K_{B\tilde A_{2}} & K_{B\tilde B_{1}} & K_{B\tilde B_2} \\
\end{array}
\right).
\end{split}
\end{align}
Then we set,
\begin{align}
\begin{split}
&K_{A\tilde A_1}=K^{0}_{A\tilde A_1}+K^{1}_{A\tilde A_1},\\&
K_{A \tilde B_1}=K^{0}_{A\tilde B_1}+K^{1}_{A\tilde B_1},\\&
K_{B \tilde A_1}=K^{0}_{B\tilde A_1}+K^{1}_{B\tilde A_1},\\&
K_{B\tilde B_1}=K^{0}_{B\tilde B_1}+K^{1}_{B\tilde B_1},\\&
K_{A\tilde A_2}=K^{0}_{A\tilde A_2}+K^{1}_{A\tilde A_2},\\&
K_{A\tilde B_2}=K^{0}_{A\tilde B_2}+K^{1}_{A\tilde B_2},\\&
K_{B\tilde A_2}=K^{0}_{B\tilde A_2}+K^{1}_{B\tilde A_2},\\&
K_{B\tilde B_2}=K^{0}_{B\tilde B_2}+K^{1}_{B\tilde B_2},
\end{split}
\end{align}
where superscript $0$ denotes the minimal ansatz value. We varies all these terms with superscript $1$ around $zero$ in some small steps  and compute the corresponding values of $S_{A\tilde A_1\tilde A_2}.$ Also first of these two  constraints in (\ref{const}) fixes all $J$'s.  The second one fixes some of the components of $L$'s. For $|A|=|B|=1$ the matrix $QR^{-1}Q^T$ is $2\times 2$ symmetric matrix.  For our case  the $L$ matrix is,
\begin{align}
\begin{split}
L=\left(
\begin{array}{cccccc}
 L_{\tilde A_{1}\tilde A_1} & L_{\tilde A_1 \tilde A_{2}} & L_{\tilde A_{1}\tilde B_{1}} & L_{\tilde A_{1} \tilde B_2} \\
 L_{\tilde A_1 \tilde A_{2}} & L_{\tilde A_2 \tilde A_{2}} & L_{\tilde A_{2} \tilde B_{1}} & L_{\tilde A_{2} \tilde B_2} \\
  L_{\tilde A_1\tilde B_{1}} & L_{ \tilde A_{2}\tilde B_1} & L_{\tilde B_{1} \tilde B_{1}} & L_{\tilde B_{1} \tilde B_2} \\
   L_{\tilde A_1\tilde B_{2}} & L_{\tilde A_{2}\tilde B_2 } & L_{ \tilde B_{1}\tilde B_{2}} & L_{\tilde B_{2} \tilde B_2} \\
\end{array}
\right).
\end{split}
\end{align}

\begin{figure}[h!]
  \centering
  \includegraphics[width=6cm]{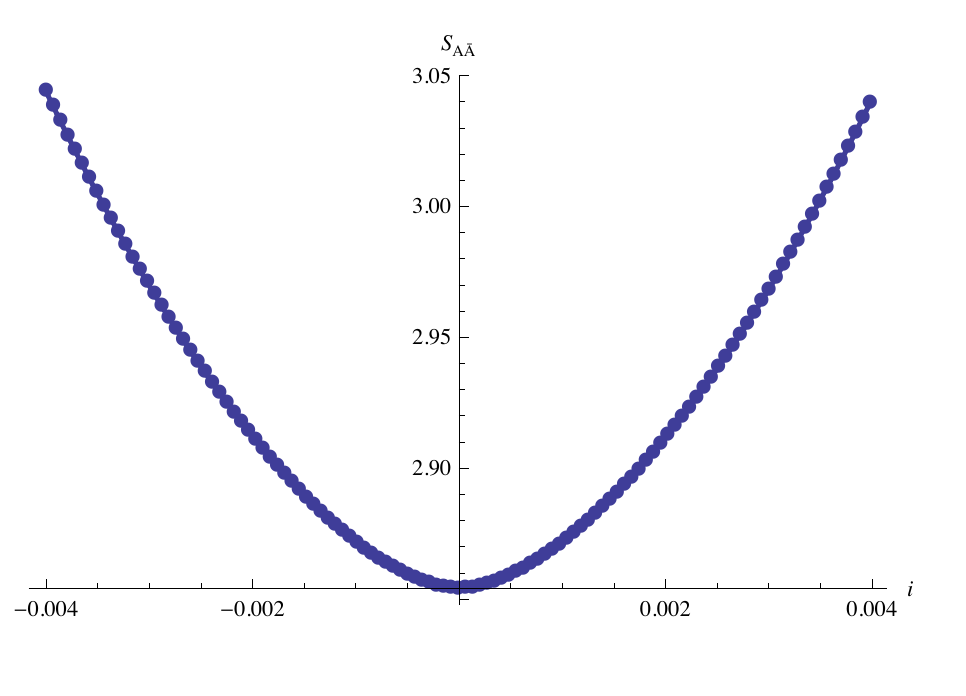}
  \caption{The variation of $S_{A\tilde A_1\tilde A_2}$ w.r.t the parameter $i$ corresponding to the bigger ansatz and its shows that it is always greater than the corresponding value coming from the minimal ansatz.}
    \label{figlarg}
    \end{figure}

This is s a symmetric matrix and hence we will have $10$ parameters.
Using the constraints, $K L^{-1}K^T=QR^{-1}Q^T$ we can determine $3$ of them. For our case we determine $L_{\tilde A_1\tilde A_1},L_{\tilde A_1\tilde B_1}$ and $L_{ \tilde B_1\tilde B_1}.$ So we have total of $15$ parameters (8 $K$'s and 7 $L$'s) and we vary them around their minimal ansatz values in some smaller steps \footnote{ We here note that inside this range sometimes it may happen that for some combinations of values of some of the parameters of the matrix $W$, where the elements are in order $A,B,\tilde A_1\tilde A_2,\tilde B_1\tilde B_2$ doesn't  remain positive definite. We exclude such combinations.}. From this we find that the  value of $S_{A\tilde A_1\tilde A_2}$ is always greater than the minimal ansatz value obtained in the previous section for all non trivial values of these extra parameters. So this shows that our minimal ansatz is good enough to produce the correct EoP.   We gave a sample plot in Fig.~(\ref{figlarg})  demonstrating this  result. We choose $d=1, N=60 , m=0.0001.$ Then we set $L_{\tilde A_2\tilde A_2}=L_{\tilde B_2\tilde B_2}=0.01, L_{\tilde A_2\tilde B_2}=10^{-7}+i$ where $i$ is the parameter. Also we set,
$K^{1}_{A\tilde A_1}=K^{1}_{A\tilde B_1}=K^{1}_{B\tilde A_1}=K^{1}_{B\tilde B_1}=0, K^{1}_{A\tilde A_2}=K^{1}_{A\tilde B_2}=K^{1}_{A\tilde A_2}=K^{1}_{A\tilde B_2}=i.$  Lastly, $L_{\tilde A_1 \tilde A_{2}} =L_{\tilde A_{1} \tilde B_2} = L_{\tilde A_{2} \tilde B_{1}}=L_{\tilde B_{1} \tilde B_2}=i.$  We vary $i$ between $-0.004$ to $0.004$ in the steps of $0.00007$ and plot the $S_{A\tilde A_1\tilde A_2}$  w.r.t $i.$ From this plot it is evident that $S_{A\tilde A_1\tilde A_2}$ is greater  than the corresponding minimal ansatz value which is $2.85393.$

Similarly we check this numerically for $|A|=1, |B|=2$ case  also and confirm that the minimal ansatz is sufficient to reproduce the correct EoP.

\section{Mutual Information}
In this section we compute various types of mutual information. As discussed in the section~(\ref{Holpuri}), these different types of the mutual information have also interesting behaviors in holographic computations. The analysis in this section will also serve as a good consistency check for our previous results of EoP. Notice the following useful relations:
\ba
 S_{A\ti{A}}&=&\frac{1}{2}I(A\ti{A}:B\ti{B})\no
&=&\frac{1}{2}I(A:B\ti{B})+\frac{1}{2}I(\ti{A}:B\ti{B}) \no
&\geq & \frac{1}{2}I(A:B)+\frac{1}{2}I(\ti{A}:\ti{B}).  \label{bdi}
\ea
Similarly we can prove
\be
S_{A\ti{A}}\geq \frac{1}{2}I(A:B)+\frac{1}{2}I(A:\ti{B}).  \label{bdj}
\ee
These suggest that if we want to minimize $S_{A\ti{A}}$ we need to
make both $I(\ti{A}:\ti{B})$ and $I(A:\ti{B})$ small.
Our holographic analysis in section~(\ref{Holpuri}) for the current setup, actually predicts
$I(\ti{A}:\ti{B})_{hol}=I(A:\ti{B})_{hol}=0$. Thus in the holographic EoP,
the minimization procedure is realized maximally. For non-holographic quantum states,
we do not expect such an extreme situation as is so in our results shown below.
We also want to mention that we confirmed the inequalities (\ref{bdi}) and (\ref{bdj}) against 
our numerical results. 

\subsection{Analysis of $I(\ti{A}:\ti{B})$}

\begin{figure}[ht]
  \centering
  \includegraphics[width=5.2cm]{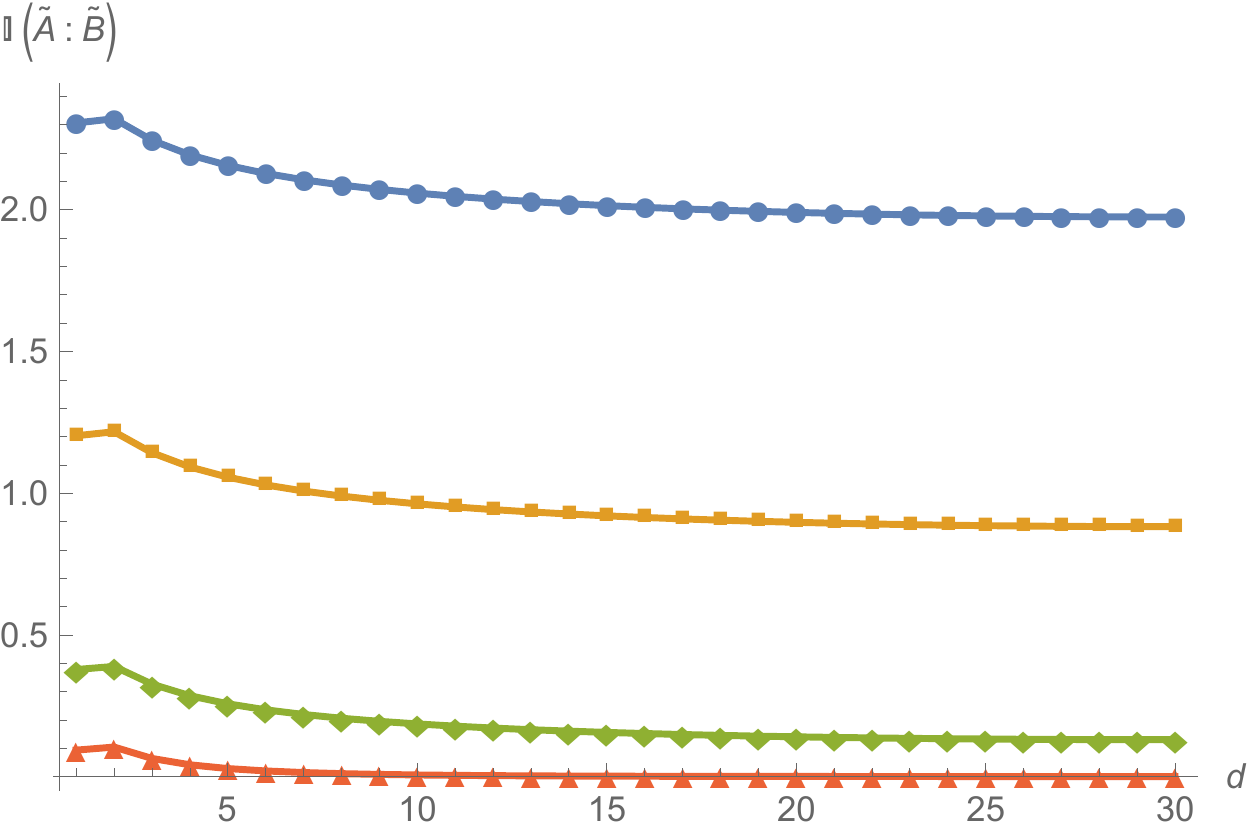}
  \includegraphics[width=5.2cm]{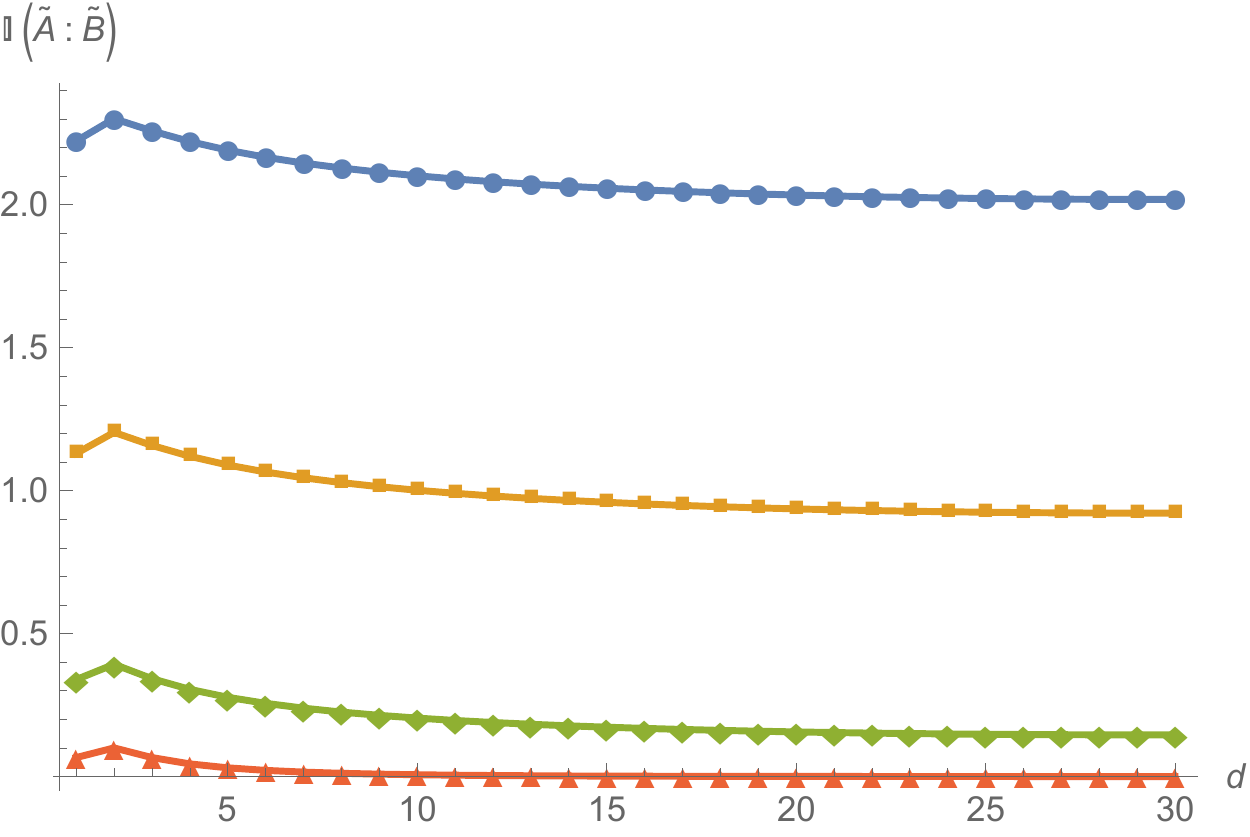}
  \includegraphics[width=5.9cm]{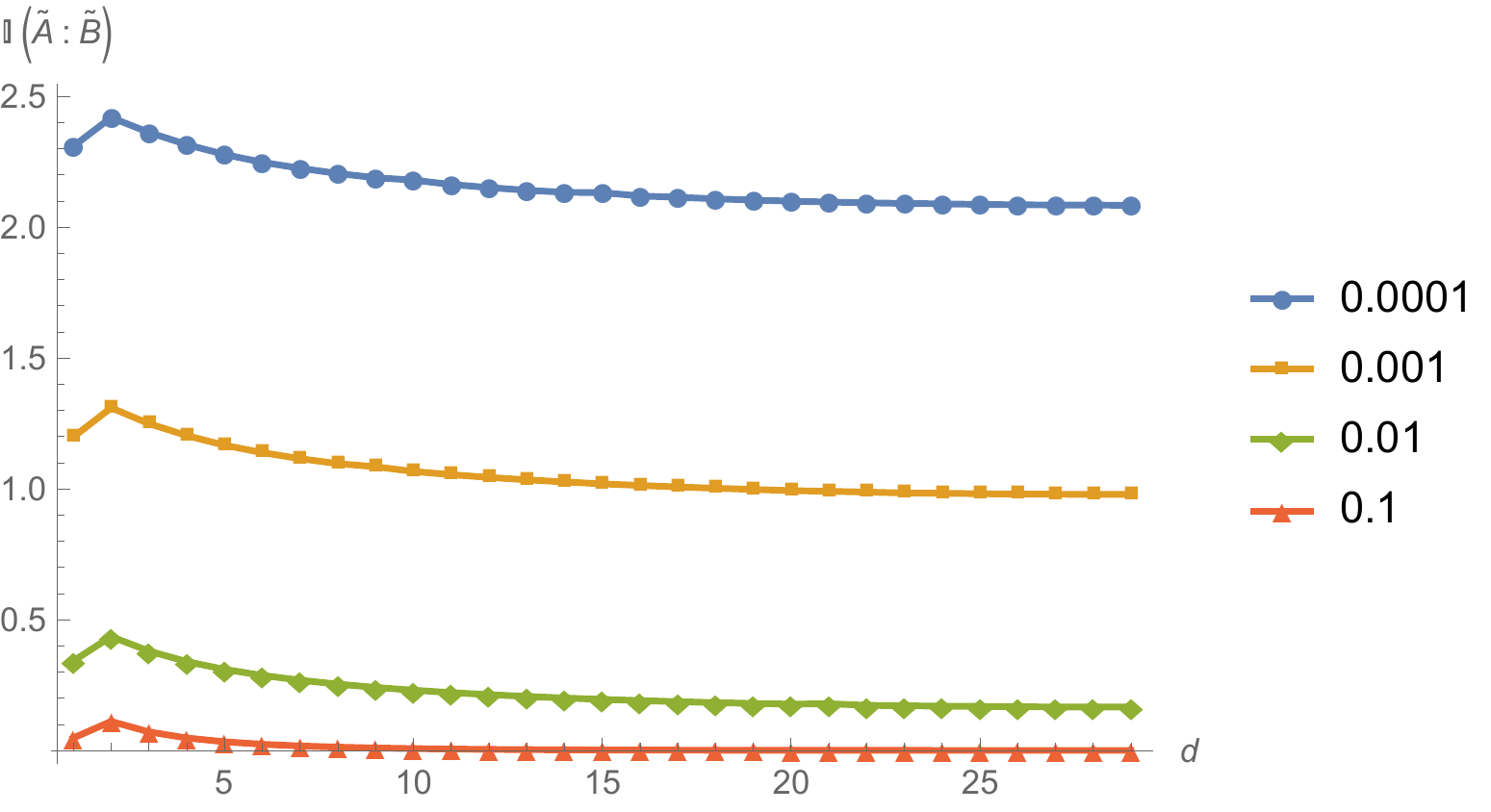}
    \caption{The left, central and the right one show $I(\ti{A}:\ti{B})$ for $|A|=|B|=1$ , $|A|=1, |B|=2$ and $|A|=|B|=2$ respectively, each for m=0.0001,0.001,0.01,0.1 indicated by different colors in each of the plots. }
\label{fig:MI}
  \end{figure}

We compute the mutual information between two subsystems $\tilde A$ and $\tilde B$ in the auxiliary Hilbert space.
We plot $I(\ti{A}:\ti{B})$ for $m=0.0001,0.001,0.01$ and $0.1$ against the the distance $d$ between $ A$ and $ B$  for $|A|=|B|=1$ , $|A|=1,|B|=2$ and $ |A|=|B|=2$ in the Fig.~(\ref{fig:MI}) using the results obtained in the previous sections.

From Fig.~(\ref{fig:MI}), it is evident that irrespective of the size of $A$ and $B$ there is a slight increase in $I(\ti{A}:\ti{B})$ around $d=2$ and then it decreases monotonically. This is consistent with our previous results for the values of parameters which minimize $S_{A\ti{A}}$ (see e.g. Fig.~(\ref{fig:A1B1X})).  For $d=2,$ there is a lattice point between $A$ and $B$. As we have argued in the section~(\ref{argument}), this enhances the correlation between $A$ and $\tilde B$ and the one between $B$ and $\tilde A$ gets enhanced and this fact gets reflected in the increase of $I(\ti{A}:\ti{B})$ around $d=2$. This further shows consistency of our results for EoP.
Non-vanishing values of $I(\ti{A}:\ti{B})$ deviates from the holographic prediction and the argument of (\ref{bdi}) implies that the minimization in our free scalar field is not as optimal as the one in holographic CFTs.

\subsection{Analysis of $I(A:\ti{B})$}
Next, we compute the mutual information between $A$ and $\tilde B$ and plot them against the distance $d$ between $A$ and $B$.

\begin{figure}[ht]
  \centering
  \includegraphics[width=5.2cm]{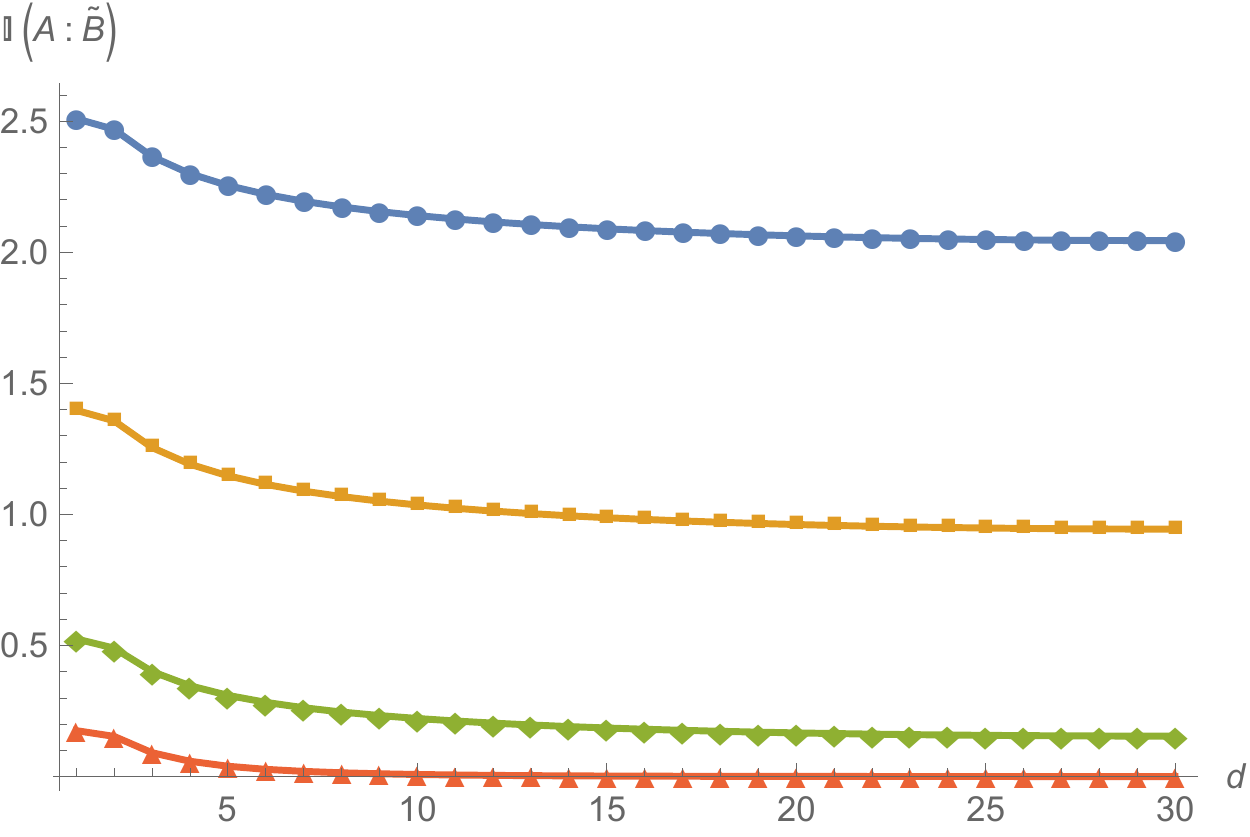}
  \includegraphics[width=5.2cm]{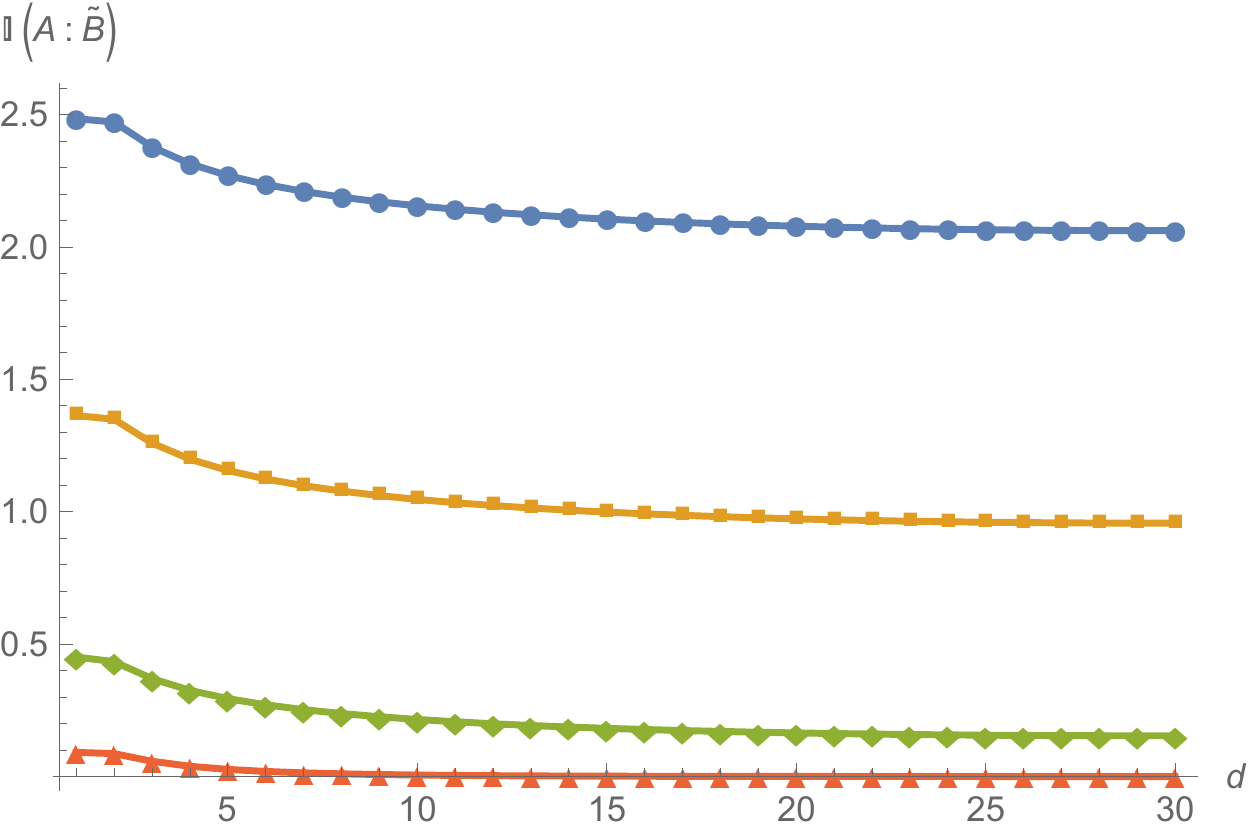}
  \includegraphics[width=5.9cm]{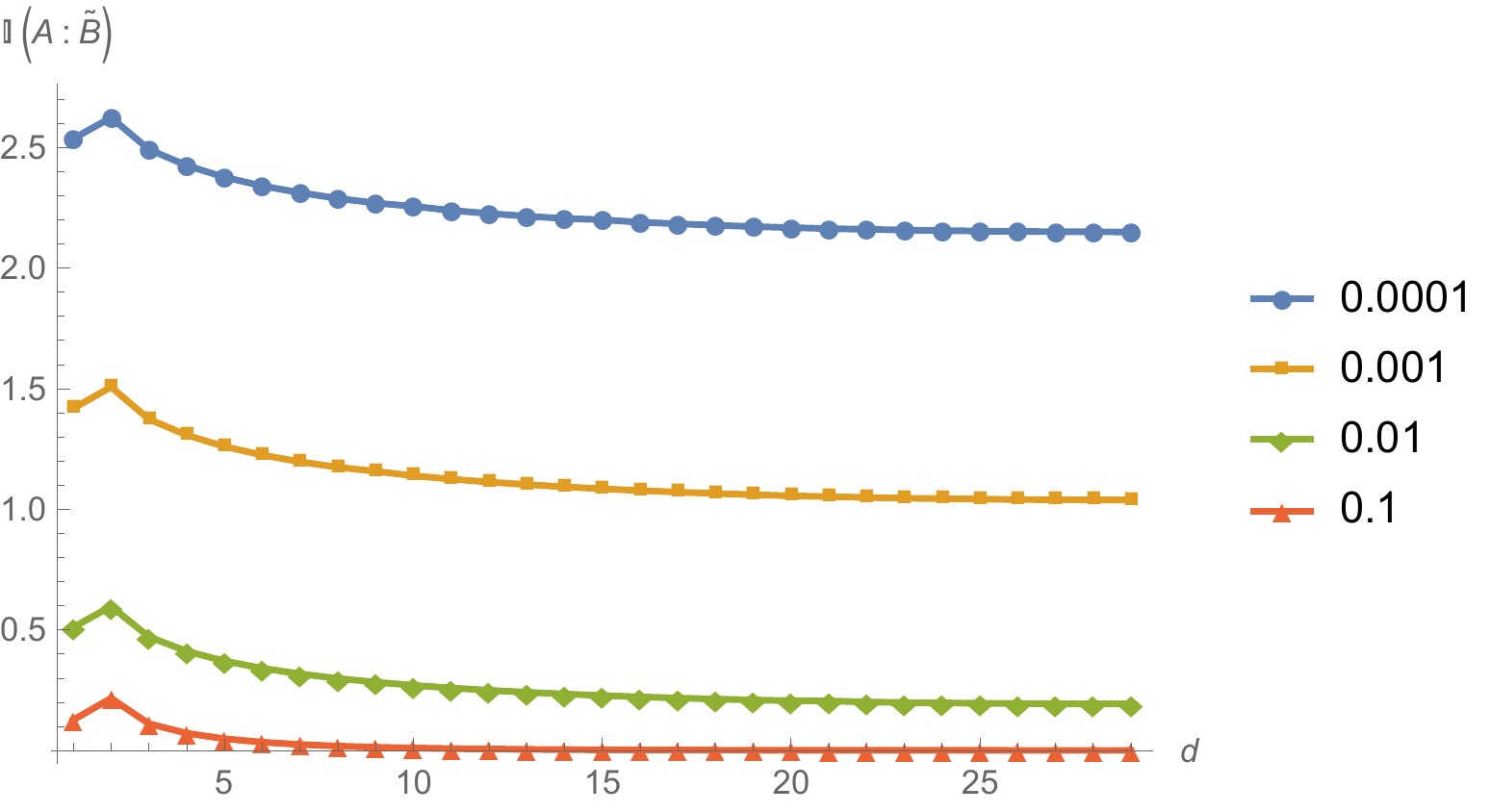}
    \caption{The left, central and the right one show $I(A:\ti{B})$ for $|A|=|B|=1$ , $|A|=1, |B|=2$ and $|A|=|B|=2$ respectively, each for m=0.0001,0.001,0.01,0.1 indicated by different colors in each of the plots. }
\label{fig:MIII}
  \end{figure}
From the Fig.~(\ref{fig:MIII}), it is evident apart from a small hump around $d=2$ $I(A:\tilde B)$ gradually decreases. We omit the details because its behavior and interpretation are very similar to the previous one $I(\ti{A}:\ti{B})$.

\subsection{Analysis of $I(A:\ti{A})$}
Lastly, we compute the mutual information between $A$ and $\tilde A.$ We demonstrate this in the plots below. Again we plot $I(A:\ti{A})$ against the the distance between $A$ and $ B$ for different mass.
\begin{figure}[ht]
  \centering
  \includegraphics[width=5.2cm]{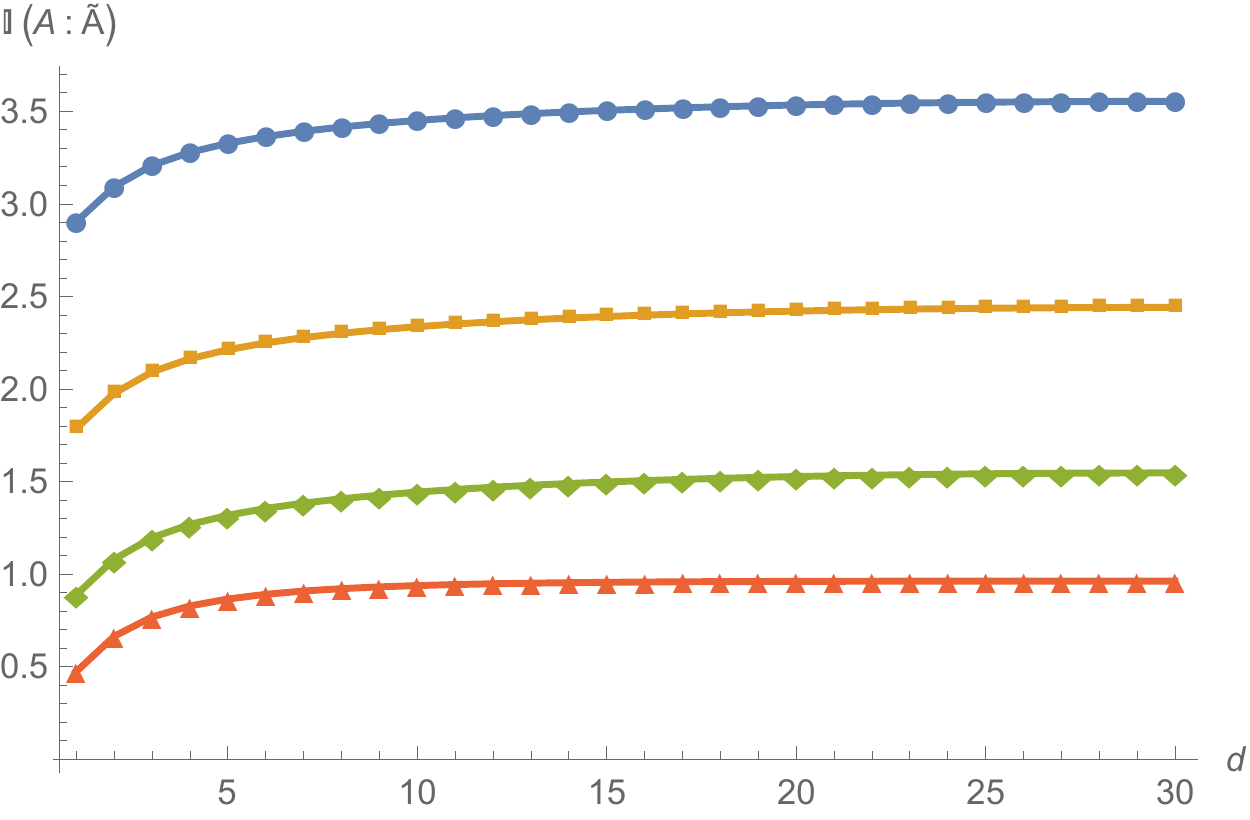}
  \includegraphics[width=5.2cm]{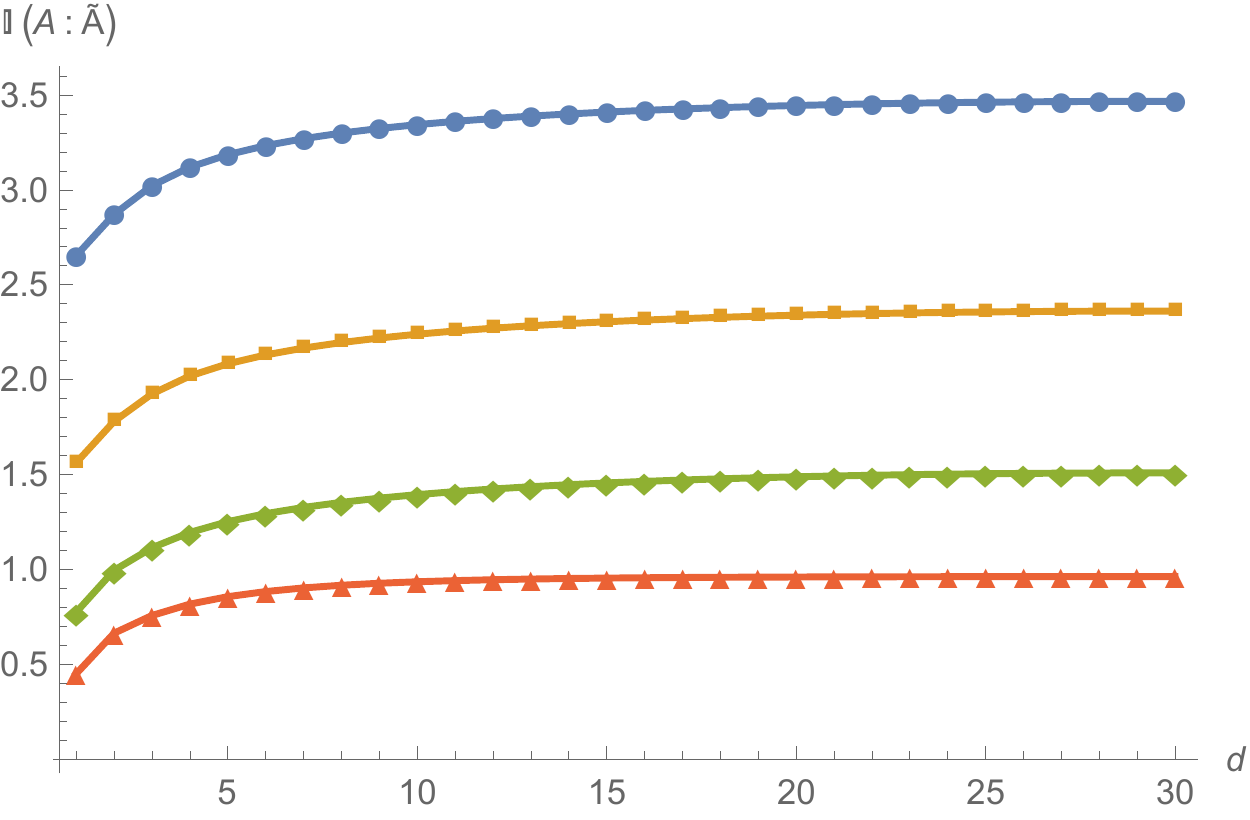}
  \includegraphics[width=5.9cm]{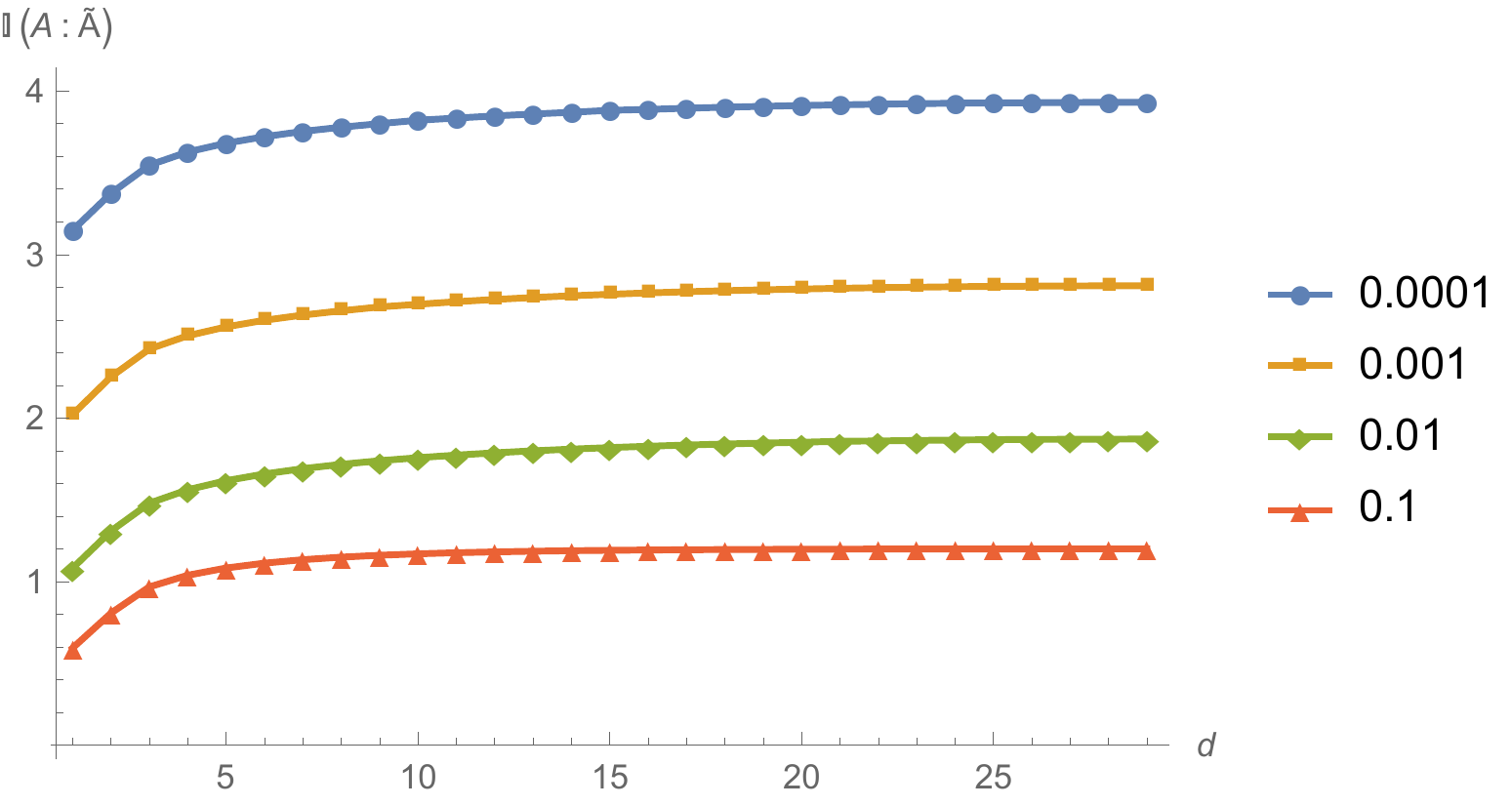}
    \caption{The left, central and the right one show $I(A:\ti{A})$ for $|A|=|B|=1$ , $|A|=1, |B|=2$ and $|A|=|B|=2$ respectively, each for m=0.0001,0.001,0.01,0.1 indicated by different colors in each of the plots. }
\label{fig:MII}
  \end{figure}
From the Fig.~(\ref{fig:MII}), it is evident that, $I(A:\ti{A})$ rather increases for all the cases unlike the previous ones. This qualitative agrees with the holographic results (see the right graph of Fig.~(\ref{fig:MIAtildeA})) if we remember that in our setup of the free scalar model,
the sizes of subsystems compete with the lattice spacing and also that we do not expect any phase transitions.

\section{Violations of Monogamy and Strong Superadditivity}\label{VioMonoSSA}

In this section we study whether the inequalities of monogamy and strong superadditivity are
satisfied or not in our numerical computations.
For a (bipartite) correlation measure $E_{\#}$ (including EoP or mutual information),
the monogamy inequality \cite{CKWmonogamy} for a tripartite state $\rho_{AB_{1}B_{2}}$
is defined by
\begin{equation}
E_{\#}(A:B_{1}B_{2})\geq E_{\#}(A:B_{1})+E_{\#}(A:B_{2}).\label{eq:Monogamy}
\end{equation}

The strong superadditivity for a 4-partite state $\rho_{A_{1}A_{2}B_{1}B_{2}}$
is defined by
\begin{equation}
E_{\#}(A_{1}A_{2}:B_{1}B_{2})\geq E_{\#}(A_{1}:B_{1})+E_{\#}(A_{2}:B_{2}).\label{eq:SSA}
\end{equation}
Note that if a measure $E_{\#}$ satisfies the monogamy for all tripartite
states, then it also satisfies the strong superadditivity. These inequalities
are regarded as desirable features of measures of quantum entanglement for mixed
states \cite{KW,SSAandMonogamy}.

It is known that the mutual information always satisfies the monogamy\footnote{
Please distinguish this from the strong subadditivity of entanglement entropy which
should be true for any quantum states. The latter is equivalent to the extensivity of mutual information. For holographic states we can derive this property from a simple geometric argument \cite{HT}.} (and thus the strong superadditivity) for holographic states \cite{HHM}. Note that the monogamy of mutual information is not satisfied for generic quantum states \cite{HHM,CHM}, for example,
in free scalar and fermion field theories.

The holographic EoP always satisfies the strong superadditivity \cite{TaUm}.
 On the other hand, it is known that the EoP does not always satisfy the strong superadditivity for generic states \cite{Ba,CW}. Therefore both monogamy of mutual information and strong superadditivity of EoP will be useful to characterize holographic states of classical
gravity duals.

\subsection{Monogamy/Polygamy}

We define the ratio of the RHS/LHS of \eqref{eq:Monogamy} as
\begin{equation}
R_{mon}=\frac{E_{\#}(A:B_{1}B_{2})}{E_{\#}(A:B_{1})+E_{\#}(A:B_{2})}.
\end{equation}
Note that the extensivity property (\ref{NIuDA}) of both EoP and mutual information tells us the ratios $R_{mon}$ is always bounded from below: $R_{mon}\geq\frac{1}{2}$. As the state gets
more quantum correlations, we expect this ratio increases. The lower bound $R_{mon}=\frac{1}{2}$
occurs when the state is classical.

\begin{figure}
\begin{centering}
\includegraphics[scale=0.65]{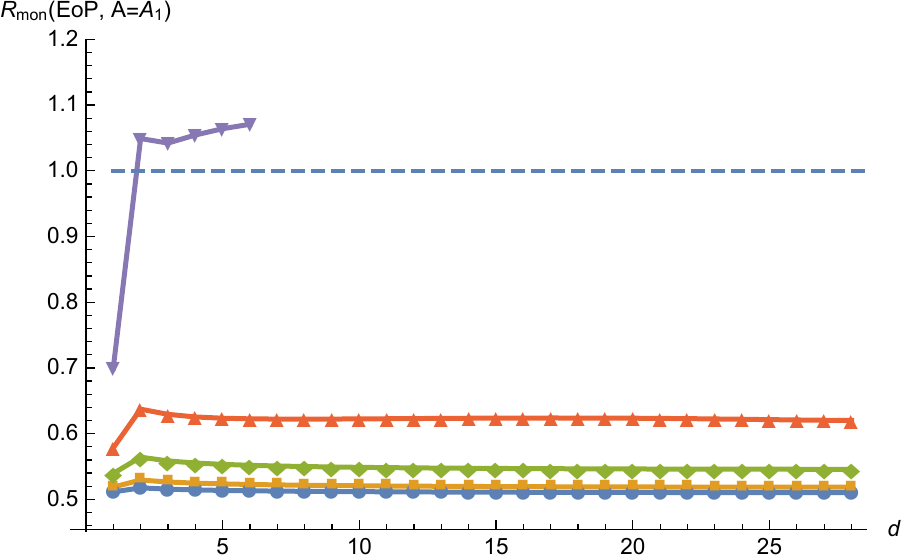}\includegraphics[scale=0.62]{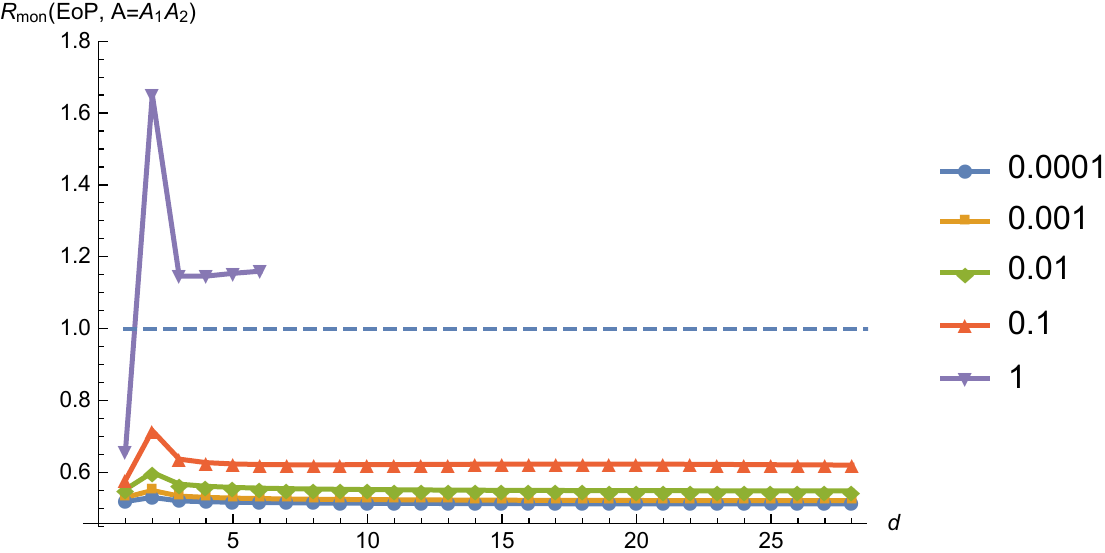}
\par\end{centering}
\centering{}\includegraphics[scale=0.65]{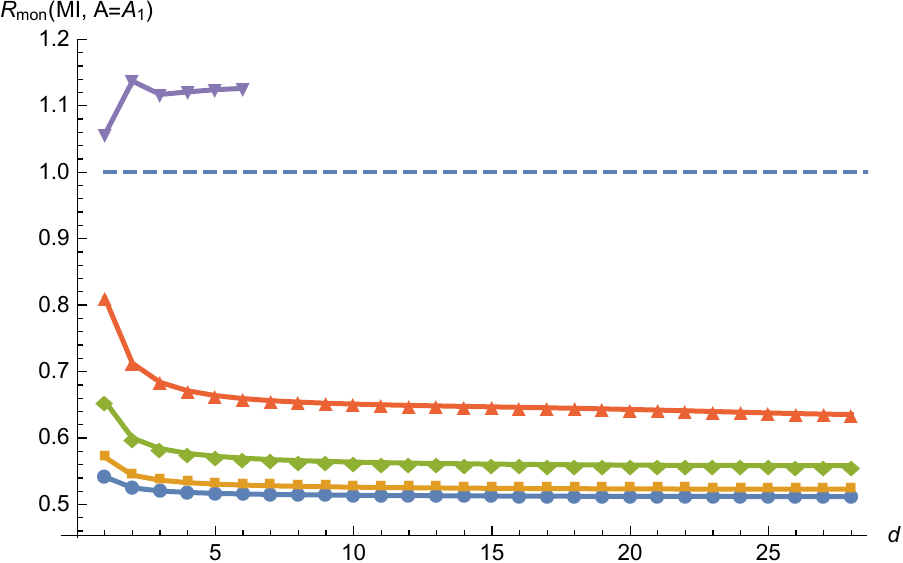}\includegraphics[scale=0.62]{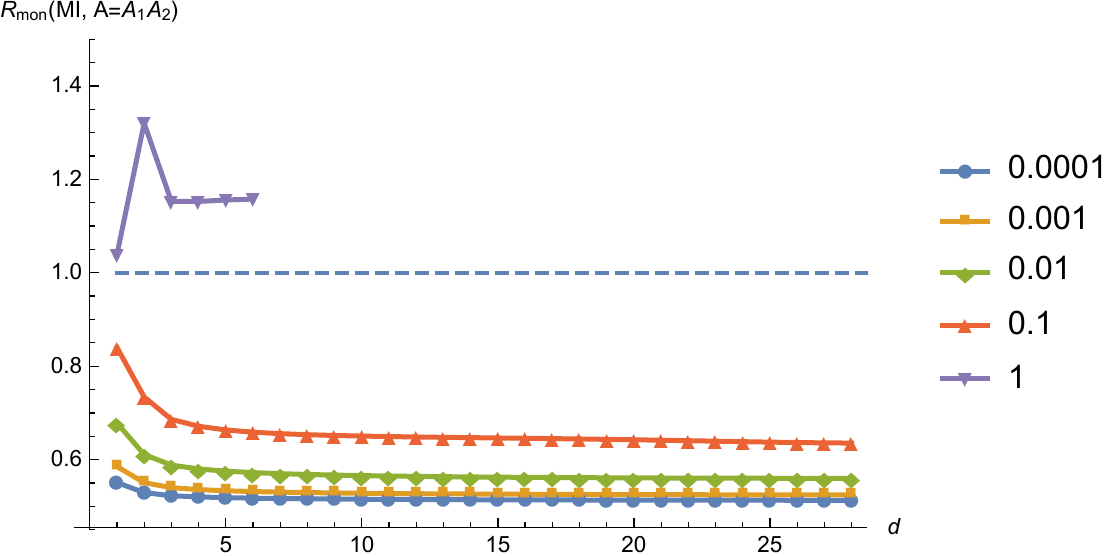}
\caption{\label{fig:Monogamy}The ratio $R_{mon}$ of the monogamy inequality
of EoP (upper graphs) or mutual information (lower graphs). 
$R_{mon}<1$ means the violation of the monogamy. For
all massless cases $m=0.1,\ 0.01,\ 0.001,\ 0.0001$ the monogamy is
violated. For massive case $m=1$ the monogamy can be satisfied. Note
that $R_{mon}$ is always bounded from below by $1/2$.}
\end{figure}

We compute this ratio $R_{mon}$ for both the EoP and mutual information (with $A=A_{1}$ or $A=A_{1}A_{2}$, while each subsystem denotes a single site as in the previous section).
We plot this in Fig.~(\ref{fig:Monogamy}) against the distance $d$ between
$A$ and $B$.

 It shows that the monogamy of EoP and that of mutual information are always violated except
the very massive case $m=1$. It can also be seen that heavier mass makes
the state more monogamous. Note that, as we discussed in the section~(\ref{Holpuri}),
EoP satisfies the polygamy rather than
the monogamy for any tripartite pure states. Hence it is natural to observe the polygamous behavior of EoP for ordinary mixed states.

For the mutual information, the difference between RHS and LHS of \eqref{eq:Monogamy}, so called the tripartite information $I_3(=RHS-LHS)$, was already computed in \cite{CHM} for a free massive scalar field theory. Their results show that as the mass goes to zero, $I_3$ gets positively divergent i.e. the monogamy is maximally violated, which is because the zero mode $\phi_0$ of the massless scalar leads to a classically maximally correlated state $\rho_{AB}\sim \int d\phi_0 |\phi_0\lb\la \phi_0|$. Our numerical results of mutual information
for our lattice free scalar model indeed reproduce the same behavior.

The EoP shows a similar behavior for large $d$ and this will be explained by the same argument of the scalar field zero mode. The new feature of EoP is that there is a peak at $d=2$ in the ratio $R_{mod}$. This will be again explained by the strong quantum correlation between $A_2$ and $B_2$ with the vacant site between $A$ and $B$, as we already observed the similar peaks in other quantities.

\subsection{Strong Superadditivity}
Next we also define the ratio of \eqref{eq:SSA} as
\begin{equation}
R_{SSA}=\frac{E_{\#}(A_{1}A_{2}:B_{1}B_{2})}{E_{\#}(A_{1}:B_{1})+E_{\#}(A_{2}:B_{2})}.
\end{equation}
 Note that the ratio has the lower bound $R_{SSA}\geq\frac{1}{2}$. The violation of
 strong superadditivity is equivalent to $R_{SSA}< 1$.

 We plot this ratio $R_{SSA}$ for the EoP and mutual information in the same manner. The results are essentially the same as that of monogamy. Except for the very massive case, the strong superadditivity is staisfied and there is a peak at $d=2$. Naturally this behavior can be interpreted as that of the monogamy violation.

\begin{figure}
\begin{centering}
\includegraphics[scale=0.65]{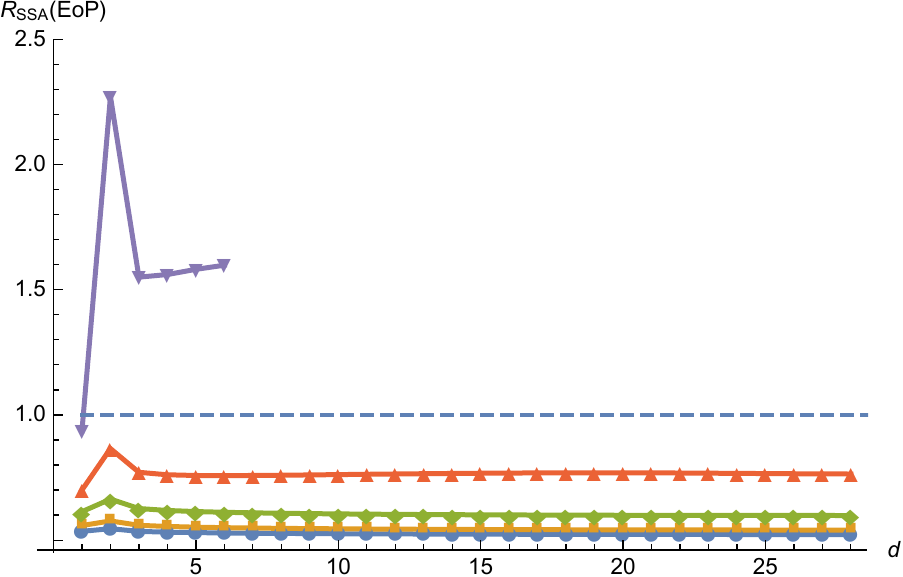}\includegraphics[scale=0.62]{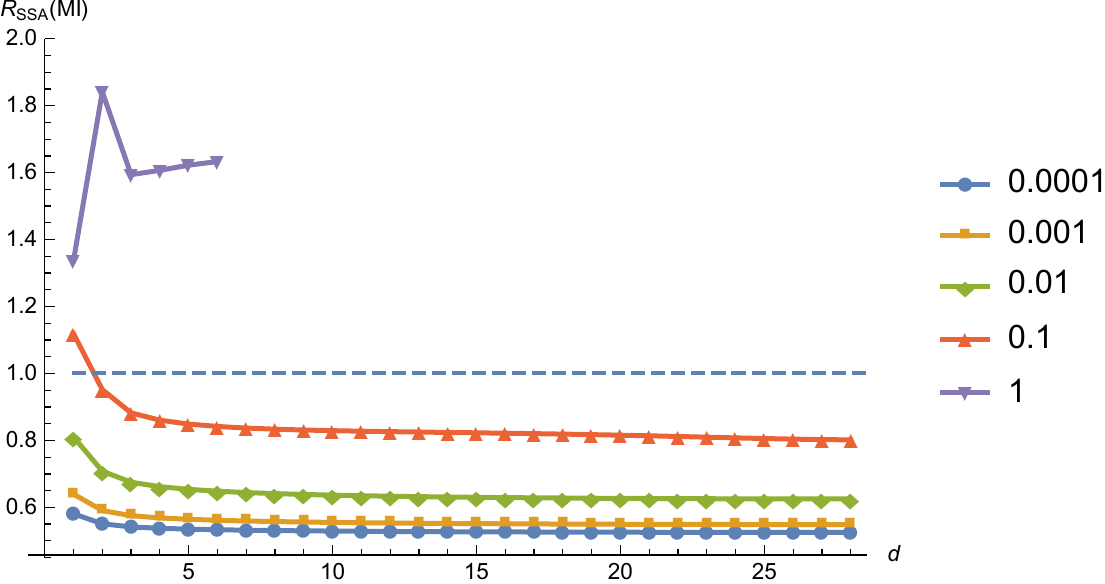}
\par\end{centering}
\centering{}\caption{\label{fig:SSA}The ratio $R_{SSA}$ of the strong superadditivity
of EoP (left) or mutual information (right). $R_{SSA}<1$ means the violation of the strong superadditivity.
For almost all massless cases $m=0.1,\ 0.01,\ 0.001,\ 0.0001$ the
strong superadditivity is violated, and for massive case $m=1$ this
can be satisfied. Note that $R_{SSA}$ is always bounded from below
by $1/2$.}
\end{figure}

\section{Conclusions and Discussions}

In this paper, we calculated the entanglement of purification (EoP) $E_P(\rho_{AB})$ for the ground state of a $1+1$ dimensional free scalar field theory. We assumed that the purified state $|\Psi\lb_{A\ti{A}B\ti{B}}$, which gives the minimum of entanglement entropy $S_{A\ti{A}}$, is described by a minimal Gaussian wave functional. Thus our numerical results at least give upper bounds of the actual EoP, defined by minimizing against all possible purifications. However, since ground states of free field theories are given by Gaussian wave functionals, there is a chance that our ansatz may provide the correct values of EoP. We presented numerical evidence that justify our minimal ansatz $|\ti{A}|=|A|$ and $|\ti{B}|=|B|$. However, we would like to leave for a future problem the final answer to the question whether our ansatz can give exact results for the EoP.

In our explicit computations, we focused on the three cases $(|A|,|B|)=(1,1),(1,2)$ and $(2,2)$ with the total lattice size $N=60$. The subsystems $A$ and $B$ can be separated by an arbitrary distance $d$ and we studied the behavior of the EoP as a function of $d$.

Our results show that the EoP is monotonically decreasing when we increase the distance $d$.
As we raise the mass of the scalar field, a power law decay is changed into an exponential
decay. These are consistent with the fact that the EoP is a measure of correlation between $A$ and $B$. We noted that the mutual information $I(A:B)$ is also monotonically decreasing as $d$ increases. However, especially in the case of $|A|=|B|=2$, we found an interesting difference between the EoP and mutual information. The EoP has a plateau for $1\leq d\leq 2$, while the mutual information does not. We argued that this plateau is qualitatively analogous to the one in the holographic EoP, which is missing for the holographic mutual information. It would be an intriguing future problem to confirm this behavior for a larger subsystem $A$, which will require more powerful numerical computations with more sophisticated numerical algorithms.

We also studied more details of our computations such as the values of parameters which specify purified states with the minimal $S_{A\ti{A}}$ and the mutual informations for other subsystems $I(\ti{A}:\ti{B})$, $I(A:\ti{B})$ and $I(A:\ti{A})$.
We noticed that they are correlated in interesting ways. In particular, some of the parameters and $I(\ti{A}:\ti{B})$ and $I(A:\ti{B})$ take maximal values at $d=2$. We argued that this behavior occurs because when there is an empty site between
$A$ and $B$, the purified site $\ti{A}$ and $\ti{B}$ are both strongly correlated with that site.
On the other hand, $I(A:\ti{A})$ is a monotonically increasing function which is similar to holographic calculations.

Moreover we examined whether the inequalities known as monogamy and strong superadditivity are satisfied or not for our numerical EoP results. It is known that in general these inequalities are not satisfied by EoP \cite{Ba,CW}. On the other hand, in the holographic EoP, the latter property turns out to be
true \cite{TaUm}. In our analysis of the free scalar field model,  we found that either of them are violated for a broad range of masses, including the massless limit. When the mass gets as large as the cut off scale, we found that both monogamy and strong superadditivity are satisfied. This behavior is similar to that for mutual information. Interestingly we observed a clear difference between them: only for EoP, there is a enhancement of monogamy and strong superadditvity
 at $d=2$. We interpreted this as the quantum correlation effect via an empty site which appears in several other quantities studied in this paper.

It would be interesting to perform similar computations based on Gaussian assumptions
for other measures, especially those quantifies quantum entanglement, such as entanglement of formation and squashed entanglement. It is also an obviously important future problem to calculate the EoP in conformal field theories directly in the continuum limit as we usually do in the replica method calculation of entanglement entropy.

\subsection*{Acknowledgements}
 We thank  Ning Bao, Pawel Caputa, Horacio Casini, Veronika Hubeny, Juan Maldacena, Xiao-liang Qi, Mukund Rangamani,  Shinsei Ryu, Noburo Shiba, Brian Swingle, Erik Tonni, Tomonori Ugajin, Kento Watanabe and Jie-qiang Wu for useful conversations. AB is supported by JSPS fellowship. AB and TT are supported by JSPS Grant-in-Aid for JSPS fellows 17F17023. TT is supported by the Simons Foundation through 
 the ``It from Qubit'' collaboration and by JSPS Grant-in-Aid for Scientific Research (A) No.16H02182. TT is also supported by World Premier International Research Center Initiative (WPI Initiative) from the Japan Ministry of Education, Culture, Sports, Science and Technology (MEXT). TT would like to thank KITP program ``Quantum Physics of Information'' where a part of this work was conducted. TT is also grateful to the conference ``20 Years Later: The Many Faces of AdS/CFT'' at Princeton and the It from Qubit workshop at Bariloche, where a part of the results of this paper were presented. AB thanks Prof. S.P.Kumar at  Department of Physics, University of Swansea  and Prof. Marika Taylor, University of Southampton for the hospitality where the final stage of this work has been completed.

\begin{appendix}




\end{appendix}


\end{document}